\documentclass[conference]{IEEEtran}
\IEEEoverridecommandlockouts
\usepackage{libertine}
\usepackage{enumitem}
\usepackage{algpseudocode}
\usepackage[ruled,linesnumbered]{algorithm2e}
\usepackage{xcolor}
\usepackage{graphicx} 
\usepackage{epstopdf}
\usepackage{subfigure}
\usepackage{amsmath,bm}
\usepackage{multirow}
\usepackage{amssymb}
\usepackage{bbding}
\usepackage{bbm}
\usepackage{marginnote}
\usepackage[hidelinks]{hyperref}

\newcommand{\zheng}{}
\newcommand{\zhengI}{}
\newcommand{\CHENG}{}
\newcommand{\zhengII}{}
\newcommand{\chengr}{}
\newcommand{\zhengIII}{}
\newcommand{\chengrr}{}
\newcommand{\zhengb}{}
\newcommand{\chengb}{}
\newcommand{\CR}{}

\newtheorem{problem}{Problem}

\DeclareMathOperator*{\argmax}{arg\,max}

\begin{document}

\title{Collectively Simplifying Trajectories in a Database: A Query Accuracy Driven Approach}

\author{Zheng Wang\IEEEauthorrefmark{1}, 
Cheng Long\IEEEauthorrefmark{1}\IEEEauthorrefmark{2}\IEEEcompsocitemizethanks{\IEEEcompsocthanksitem\IEEEauthorrefmark{2}Cheng Long is the corresponding author.},
Gao Cong\IEEEauthorrefmark{1},
Christian S. Jensen\IEEEauthorrefmark{3}\\
\IEEEauthorrefmark{1}School of Computer Science and Engineering, Nanyang Technological University, Singapore\\
\IEEEauthorrefmark{3}Department of Computer Science, Aalborg University, Denmark\\
zheng011@e.ntu.edu.sg, \{c.long,gaocong\}@ntu.edu.sg, csj@cs.aau.dk
}

\maketitle

\pagenumbering{arabic}
\setcounter{page}{1}

\begin{abstract}

Increasing and massive volumes of trajectory data are being accumulated that may serve a variety of applications, such as mining popular routes or identifying ridesharing candidates. As storing and querying massive trajectory data is costly, trajectory simplification techniques have been introduced that intuitively aim to reduce the sizes of trajectories, thus reducing storage and speeding up querying, while preserving as much information as possible. Existing techniques rely mainly on {\chengr hand-crafted} error measures when {\zheng{deciding which point to drop 
when
simplifying a trajectory.}} While the hope may be that such simplification affects the subsequent usability of the data only minimally, the usability of the simplified data remains largely unexplored. Instead of using error measures that indirectly may to some extent yield simplified trajectories with high usability, we adopt a direct approach to simplification and present the first study of query accuracy driven trajectory simplification, where the direct objective is to achieve a simplified trajectory database that preserves the query accuracy of the original database as much as possible. {\zheng{Specifically, we propose a multi-agent reinforcement learning based solution 
{\CHENG with two agents} working cooperatively {\CHENG to collectively simplify trajectories in a database} while optimizing query usability.
}}
Extensive experiments on four real-world trajectory datasets show that the solution is capable of consistently outperforming baseline solutions over various query types and dynamics.

\end{abstract}


\section{INTRODUCTION}
\label{sec:intro}

A trajectory is a sequence of time-stamped locations that describe the movement of an object over time. Massive amounts of trajectory data are being accumulated and used in diverse applications, such as discovering popular or anomalous routes in a city~\cite{chen2011discovering,zhang2023online}, analyzing animal migration patterns~\cite{li2011movemine}, and performing sports analytics~\cite{wang2019effective,wang2021similar,zhang2022predicting}. 
The accumulation of trajectory data introduces at least two challenges
~\cite{lin2021error,zhang2018trajectory}: (1) storing the data is expensive, and (2) querying the data is time-consuming. These challenges can be addressed by conducting trajectory simplification, which aims to drop points from trajectories to save the storage cost and speed up query processing. 
The rationale is that not all points in a trajectory carry equally important information, so that dropping unimportant ones may be acceptable. For example, if the location of an object is sampled regularly and the object does not move for a while then only the first and last positions during the period of inactivity are important, and those in-between may be dropped without loosing information.
Next, the efficiency of query processing is improved, at the expense of query results becoming approximate.

Indeed, the extent to which a collection of simplified trajectories enables accurate query results has been used widely as a measure of the quality of a simplification technique in empirical studies of trajectory simplification~\cite{cao2006spatio,lin2021error,zhang2018trajectory}. 
For example, Zhang et al.~\cite{zhang2018trajectory} evaluate existing simplification techniques in terms of their ability to produce simplified trajectories that affect the accuracy of range, $k$NN, and join queries as well as clustering minimally. 
%
While there are many proposals for trajectory simplification~\cite{hershberger1992speeding,marteau2009speeding,long2014trajectory,wang2021trajectory,potamias2006sampling,muckell2011squish,muckell2014compression}, they all assume a storage budget and aim to produce simplified trajectories that minimize the difference from the original trajectories according to a given difference notion.
No proposals aim to optimize directly the query accuracy offered by the simplified trajectories.
%
We call this line of study \emph{Error-Driven Trajectory Simplification (EDTS)}. 




In addition, existing simplification techniques are local in nature and operate on a per-trajectory basis as opposed to being global in nature and operating on a database of trajectories. Specifically, they aim to simplify a given trajectory $T$ within a budget $r\cdot |T|$, where {\chengr $r\in (0, 1]$} is the compression ratio. When these techniques are used to simplify a database of trajectories according to a compression ratio $r$, they simplify each trajectory in the database \emph{separately} according to compression ratio $r$. This is likely sub-optimal in cases where trajectories have different sampling rates or different complexities. Intuitively, trajectories with higher sampling rates or lower complexity are candidates for simplification with larger compression ratios.

Therefore, we propose a new trajectory simplification problem, called \emph{Query accuracy Driven Trajectory Simplification (QDTS)}. Given a trajectory database $D$ and a storage budget, the problem is to find a simplified trajectory database $D'$ that preserves the accuracy of query results as much as possible, compared to the query results on $D$. 
QDTS differs from the existing EDTS problem. First, it considers a different objective of trajectory simplification, namely that of preserving query accuracy directly, rather than through minimizing an error measure, as in the EDTS problem. Second, it is global in nature and takes a trajectory database as input and outputs a simplified trajectory database that satisfies a specified storage budget as a whole, 
instead of simplifying each trajectory in isolation according to a budget, as do existing EDTS techniques~\cite{hershberger1992speeding,marteau2009speeding,wang2021error}.

\noindent{\textbf{Challenges.}}
An immediate solution to the QDTS problem is to reduce it (i.e., a \emph{database-level} simplification problem) to a \emph{trajectory-level} problem by applying simplification to each trajectory \emph{separately} with a proportional budget. Specifically, 
it simplifies each trajectory $T$ with the budget of $r\cdot |T|$.
It is obvious that the resulting simplified database would contain $M$ simplified trajectories and have at most $r\cdot \sum_{T\in D}|T| = r\cdot N$ points,
{\chengr where $N$ denotes the total number points in the database}. While this solution needs minimal design efforts, it {\CHENG suffers from two main issues}. \underline{Issue 1: Uniform compression ratio}: It applies the same compression ratio to each trajectory, which would be sub-optimal for trajectories with different sampling rates and/or different complexities. 
\underline{Issue 2: Query accuracy unawareness}: Existing algorithms~\cite{hershberger1992speeding,marteau2009speeding,long2014trajectory,wang2021trajectory,potamias2006sampling,muckell2011squish,muckell2014compression} (all of which operate at trajectory-level) aim to optimize some form of error metric that quantifies the difference between an original and a simplified trajectory.
As a consequence, simply using any of these algorithms 
{\CHENG cannot help to} optimize directly the query accuracy of the simplified database.

Another solution is to consider all trajectories in the database $D$ \emph{collectively} during the course of simplification. Consider a top-down approach, in which we start with the most simplified database, i.e., each simplified trajectory $T'$ of an original trajectory $T$ contains only the first and last points of $T$. We then iteratively introduce points from the original database according to some selection criterion until the budget is exhausted. 
Alternatively, we can adopt a bottom-up approach, in which we start from $D$ and iteratively drop points until {\chengr the remaining points are within the budget}. This solution considers all trajectories \emph{collectively} {\CHENG when simplifying trajectories, enabling different trajectories to be simplified with different compression ratios depending on their complexities.} {\CHENG Therefore,} it avoids the first issue mentioned above. {\CHENG However}, this solution still does not contend with the second issue. Furthermore, the solution operates at the database level, whose scale is typically much larger than that of a single trajectory. For example, the Geolife dataset used in our experiment contains millions of points, while individual trajectories contain only around one thousand points. This brings up a third issue. \underline{Issue 3: Lack of scalability}: Operating at the database level, the solution needs to repeatedly choose a point from among a very large set of points.

%
\noindent\textbf{New Solution.} Motivated by the above discussion, we propose a new solution called \texttt{RL4QDTS} for trajectory database simplification, which avoids all the three issues. {\CHENG \texttt{RL4QDTS} has two core ideas. \underline{First}, it considers all trajectories in the database \emph{collectively} for simplification. Specifically, it starts with the most simplified database, 
then introduces original points into the simplified database iteratively until its budget is exhausted. For better efficiency, it uses an index that partitions the database into sub-spaces, called \emph{spatio-temporal cubes}. Whenever it needs to choose a point to introduce into the database, it first chooses a cube based on the index and then chooses a point in the cube.} 
%
%
{\CHENG To partition a database of trajectories}, one immediate idea is to partition {\CHENG along the spatial and temporal dimensions} with a predefined granularity, e.g., setting a grid size for the spatial {\CHENG dimensions} and a time duration for the temporal {\CHENG dimension}. Nevertheless, the granularity is hard to set appropriately and is unlikely to work across databases. 
{\CHENG Small} cubes (corresponding to a {\CHENG fine} granularity) contain few candidate points,
{\CHENG making it difficult to find good points to introduce.}
{\CHENG Large} cubes contain many points, {\CHENG making it costly to choose one point within a cube.}
{\CHENG Therefore, {\CHENG \texttt{RL4QDTS} builds} an octree (a three-dimensional variant of the quadtree for spatio-temporal points) to partition the database into cubes. The octree}
provides different resolutions of data cubes organized in a tree structure, {\CHENG making it possible to} choose cubes with different sizes {\CHENG flexibly and adaptively} by traversing the tree from the root node to {\CHENG an appropriate node}. 
{\chengrr We adopt the octree for its simplicity and leave other indexes, e.g., kd-tree~\cite{bentley1975multidimensional}, for future exploration.}

\underline{Second}, 
\texttt{RL4QDTS} leverages multi-agent reinforcement learning
{\CHENG to choose a point iteratively such that the query accuracy based on the simplified database involving the chosen points is optimized.}
%
%
Specifically, it employs an agent (called Agent-Cube) to find an octree node with a cube of an appropriate size. 
{\CHENG Then, it employs} another agent (called Agent-Point) to choose a point {\CHENG in the cube chosen by Agent-Cube} and introduces the point into the simplified database. 
{\CHENG Specifically, the two agents employ Markov decision processes (MDP)~\cite{sutton2018reinforcement} that are designed such that the two agents optimize cooperatively the query accuracy on the simplified database.
}
%
%
{\CHENG The \texttt{RL4QDTS} algorithm then leverages the learned polices of the two agents for simplifying a trajectory database.}
{\CHENG In summary, the first idea enables a solution that avoids the \underline{first and third issues}, and the second idea 
is to address the \underline{second issue}.}

Overall, we make the following contributions. 
\begin{itemize}
    \item We propose the QDTS problem that aims to find a simplified trajectory database within a given storage budget that preserves the query accuracy on the simplified database as much as possible. This is the first systematic study of this line of trajectory simplification. (Section~\ref{sec:problem})
    \item We develop a multi-agent reinforcement learning based solution called \texttt{RL4QDTS}. 
    {\CHENG It simplifies a database of trajectories collectively with the aim of optimizing the query accuracy on the simplified database, while leveraging an index on the trajectory data for better efficiency.} 
    We show that the objective of \texttt{RL4QDTS} {\CHENG is well aligned with that of} the QDTS problem.
    (Section~\ref{sec:method})
    \item We conduct experiments on {\chengrr four real-world trajectory datasets}, {\chengb showing that} \texttt{RL4QDTS} consistently outperforms existing EDTS solutions across \emph{two} types of adaptions, \emph{four} error measures, and \emph{varying} storage budgets for \emph{five} query operators. {\zhengI{For example, it achieves the improvement of up to 35\% for range query, 41\% and 28\% for two kinds of $k$NN query, 35\% for similarity query and 40\% for clustering than the best baselines.}}
    (Section~\ref{sec:experiment})
\end{itemize}



\section{RELATED WORK}
\label{sec:related}

\smallskip
\noindent\textbf{Error-Driven Trajectory Simplification.} The error-driven trajectory simplification aims to simplify a trajectory within a given storage budget and to minimize an
error measure of the simplified trajectory.
Many studies have been conducted on this problem, among which some focus on the batch mode (where full access to a trajectory is attained throughout the process)~\cite{hershberger1992speeding,marteau2009speeding,long2014trajectory,wang2021trajectory} and others on the online mode (where a trajectory is inputted in an online fashion and those points that {\chengr have} been dropped are no longer accessible)~\cite{potamias2006sampling,muckell2011squish,muckell2014compression}.
We review the studies on the batch mode, which is the focus of this paper, as follows.
Specifically, Top-Down~\cite{hershberger1992speeding} adapts the traditional Douglas-Peucker algorithm~\cite{douglas1973algorithms}. 
The algorithm starts with two points (the first and last) of a trajectory.
Then, it repeatedly inserts a point with the largest error until the size of the simplified trajectory reaches the storage budget.
Bottom-Up~\cite{marteau2009speeding} adopts a reverse strategy. It 
scans all points of the input trajectory and repeatedly drops the point with the smallest error until the number of remaining points is within the storage budget. Long et al.~\cite{long2014trajectory} propose Span-Search, which is designed specifically to preserve direction information in trajectory simplification.
Recently, Wang et al.~\cite{wang2021trajectory} propose a reinforcement learning based method called RLTS+ for trajectory simplification. It adopts the Bottom-Up strategy and drops points based on a learned policy instead of using the heuristic rules seen in previous studies.

Overall, the above studies aim to minimize a given error measure while simplifying a trajectory. 
However, they largely disregard data usability as an objective of simplification algorithms. 
As a matter of fact, one of the main motivations for simplification 
is to improve query efficiency. Consequently, data usability should be treated as a key factor to indicate the quality of trajectory simplification. 
Indeed, data usability has been used to compare existing simplification algorithms in several empirical studies~\cite{cao2006spatio,zhang2018trajectory,lin2021error}. 
An early study~\cite{cao2006spatio} evaluates several error measures in trajectory simplification
and analyzes the soundness of the measures for queries. Zhang et al.~\cite{zhang2018trajectory} consider four spatio-temporal queries (i.e., range query, $k$NN query, join query, and clustering) on a trajectory database and design the corresponding measures to evaluate the quality of existing trajectory simplification algorithms. A recent evaluation study~\cite{lin2021error} verifies the query qualities {\CHENG of error-bounded} trajectory simplification algorithms~\cite{lin2017one,lin2019one,liu2015bounded,meratnia2004spatiotemporal} that simplify a trajectory with a given error tolerance and aim to minimize {\CHENG the size of} a simplified trajectory. 
However, data usability is only used as evaluation measures in these studies~\cite{cao2006spatio,zhang2018trajectory,lin2021error} to understand how well the existing simplification algorithms support various types of queries, {\chengr but} not considered in simplification algorithm design. 
{\zhengb We note that existing optimal algorithms~\cite{chan1996approximation} for EDTS problem cannot be applied or adapted to the QDTS problem studied in this work. They are usually dynamic programming based or binary search based and have high time costs (i.e., cubic time complexity for quite a few error measures), and thus they are not practical.}
%
%





\smallskip
\noindent
\textbf{Other Types of Trajectory Simplification.} 
Other studies of trajectory simplification include: (1) studies simplifying a trajectory such that the error of the simplified trajectory is bounded and as many points as possible are dropped, considering batch mode~\cite{long2013direction, hershberger1992speeding,keogh2001online,muckell2014compression} and online mode~\cite{wang2021error,lin2017one,liu2015bounded,meratnia2004spatiotemporal},
(2) studies simplifying a trajectory by immediately deciding whether to keep or drop an incoming point (also called \emph{dead reckoning})~\cite{potamias2006sampling,muckell2011squish,muckell2014compression}, 
{\zhengIII{(3) {\chengrr a study which} develops a trajectory quantization method {\chengrr that} assigns a smaller number of bits for each trajectory point - {\chengrr when} it assigns 0 bits to a point, {\chengrr it means to} drop the point~\cite{wang2020ppq}}},  {\zhengb and (4) a study which develops optimal algorithms for the curve simplification under different settings of distance/error measures and restrictions of points to be kept~\cite{van2018global}.}
These studies {\CHENG do not return trajectories with sizes bounded by user-specified parameters}
and thus cannot be used for our QDTS.

\smallskip
\noindent
\textbf{{\CHENG Road Network-based} Trajectory Compression.} 
{\CHENG Road network-based trajectory compression~\cite{han2017compress,song2014press,li2020compression,li2021trace} aims to compress trajectories that are generated by objects in road networks}.
Specifically, raw trajectories are initially map-matched to an underlying road network to obtain map-matched trajectories that consist of sequences of road segments. The map-matched trajectories are treated as strings, where each road segment is considered as a character. Then, string compression algorithms such as Huffman coding~\cite{huffman1952method} can be utilized to compress the trajectories with or without information loss. 
In contrast, {\CHENG we aim to reduce trajectory data in its original form, i.e., as a sequence of time-stamped locations, without the input of a road network.}

\smallskip
\noindent
\textbf{Reinforcement Learning.} Reinforcement Learning (RL) aims 
to guide agents on how to take actions to maximize a cumulative reward in an environment, where the environment is usually modeled as a Markov decision process (MDP), involving states, actions, and rewards~\cite{sutton2018reinforcement}. Recently, RL has been applied successfully to solve many algorithmic problems, such as similarity search~\cite{wang2020efficient}, index learning~\cite{yang2020qd}, fleet management~\cite{lin2018efficient}, {\CHENG and trajectory simplification~\cite{wang2021trajectory, wang2021error}}. 
{\CHENG Our study differs from the existing studies of RL-based trajectory simplification in three aspects.}
1) Our method simplifies trajectories collectively in a database, rather than simplifying each trajectory with an uniform compression ratio~\cite{wang2021trajectory}. 2) Our method optimizes the query accuracy on a simplified database, rather than minimizing an error measure on a single trajectory~\cite{wang2021trajectory} {\CHENG or minimizing the size of a simplified trajectory~\cite{wang2021error}. Both existing methods are query un-aware. 3) 
Our method builds an octree on the trajectory database and iteratively chooses a point by choosing \emph{a cube} {\chengr with a traversal on an octree} and then choosing \emph{a point} within the cube. It leverages two agents for the two decision making processes of choosing a cube and a point. These decision making processes are different from those in the existing methods~\cite{wang2021trajectory,wang2021error} and the corresponding designs (e.g., those of MDPs) are different.}
%
\section{PRELIMINARIES AND PROBLEM STATEMENT}
\label{sec:problem}

\subsection{Preliminaries}
\label{subsec:pre}

\textbf{Trajectories and Segments.} 
A \emph{trajectory} $T$ is a sequence of {\CHENG time-stamped} points: $T=\langle p_1, p_2, ..., p_n \rangle$, where $n$ is the length of $T$ (i.e., $n=|T|$). Each point $p_i$ $(1 \leq i \leq n)$ is a triple $p_i=(x_i, y_i, t_i)$, {\CHENG indicating that the moving object is} at location $(x_i, y_i)$ at time $t_i$. 
{\zhengI{We define the line $\overline{p_ip_{i+1}}$ linking two neighboring 
{\CHENG points}
as a segment in the trajectory.}}
{\CHENG Thus, the trajectory $T$ corresponds to a sequence of segments $\overline{p_1p_{2}}, \overline{p_2p_{3}}, ..., \overline{p_{n-1}p_{n}}$.}
%
%
A \emph{trajectory database} $D$ consists of a set of trajectories.
We define $N$ to be the total number of points in $D$.


\smallskip
\noindent
\textbf{{\CHENG Trajectory} Simplification and Errors.} \emph{Trajectory simplification} aims to eliminate points from a trajectory $T$ to obtain a simplified trajectory $T'$ of the form $T' = \langle p_{s_1},p_{s_2},...,p_{s_m} \rangle$, where $m \leq n$ and $1=s_1 < s_2 ... < s_m=n$. 
{\zhengI{We similarly call the line $\overline{p_{s_i}p_{s_{i+1}}}$ linking two neighbouring points in the simplified trajectory as a simplified segment.
%
The simplified trajectory $T'$ indicates that the object moves along a simplified segment $\overline{p_{s_j}p_{s_{j+1}}}$ ($1 \leq j \leq m-1$), which approximates the movement along a sequence of segments $\overline{p_{s_j}p_{s_j+1}}, \overline{p_{s_j+1}p_{s_j+2}},..., \overline{p_{s_{j+1}-1}p_{s_{j+1}}}$ as indicated by the original trajectory $T$. 
Thus, we call the simplified segment $\overline{p_{s_j}p_{s_{j+1}}}$ an \emph{anchor segment} for each of the points $p_{s_j}, p_{s_j+1},..,p_{s_{j+1}-1}$.
}}

\begin{figure}
  \centering
  \includegraphics[width=6cm]{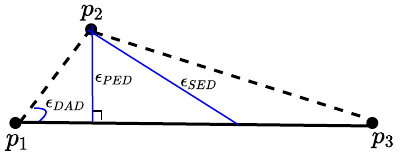}\\
  \setlength{\belowcaptionskip}{1pt}
  \vspace{-6mm}
  \caption{Error measurements.}\label{fig:error_measurement}
  \vspace{-6mm}
\end{figure}

%
To measure the information loss of the 
simplification, {\CHENG several} \emph{error measures} have been proposed, 
including Synchronized Euclidean Distance (SED)~\cite{potamias2006sampling, muckell2011squish, muckell2014compression, meratnia2004spatiotemporal}, Perpendicular Euclidean Distance (PED)~\cite{meratnia2004spatiotemporal, liu2015bounded, liu2016novel, bellman1961approximation}, Direction-aware Distance (DAD)~\cite{ke2016online, ke2017efficient, long2014trajectory, long2013direction}, and Speed-aware Distance (SAD)~\cite{muckell2014compression}.
%
These measures are defined in two steps.
First, the error of a simplified segment {\CHENG $\overline{p_{s_j}p_{s_{j+1}}}$} ({\CHENG denoted by} $\epsilon(\overline{p_{s_j}p_{s_{j+1}}})$) 
is defined as the maximum error of {\CHENG an original point $p_i$ that takes the segment as its anchor segment (denoted by $\epsilon(\overline{p_{s_j}p_{s_{j+1}}} | p_i)$)}:
\begin{equation}
    \epsilon(\overline{p_{s_j}p_{s_{j+1}}}) = \max_{s_j \le i < s_{j+1}}\epsilon(\overline{p_{s_j}p_{s_{j+1}}} | p_i),
    \label{equ:error-segment}
\end{equation}
where $\epsilon(\overline{p_{s_j}p_{s_{j+1}}} | p_i)$ can be instantiated with SED, PED, DAD, or SAD. Figure~\ref{fig:error_measurement} illustrates $\epsilon_{SED}(\overline{p_1p_3} | p_2)$, $\epsilon_{PED}(\overline{p_1p_3} | p_2)$, and 
$\epsilon_{DAD}(\overline{p_1p_3} | p_2)$. Detailed definitions can be found in an evaluation paper~\cite{zhang2018trajectory}.
Second, the error of the simplified trajectory $T'$ ({\chengr denoted by} $\epsilon(T')$) is defined as the maximum error of its simplified segments:
\begin{equation}
    \epsilon(T') = \max_{1\le j\le m-1}\epsilon(\overline{p_{s_j}p_{s_{j+1}}})
    \label{equ:error}
\end{equation}

\smallskip
\noindent
\textbf{Error-Driven Trajectory Simplification (EDTS).} In the EDTS problem~\cite{hershberger1992speeding,marteau2009speeding,wang2021trajectory,long2014trajectory}, given a trajectory $T=\langle p_1, p_2, ..., p_n \rangle$ and a storage budget 
$W$, it aims to find a simplified trajectory $T' = \langle p_{s_1},p_{s_2},...,p_{s_m} \rangle$ such that $|T'| \leq W$ and $\epsilon(T')$ is minimized, where $\epsilon(T')$ is SED, PED, DAD, or SAD.


\subsection{Problem Definition}
\label{subsec:problem}
We define a new problem, called Query Accuracy Driven Trajectory Simplification (QDTS), which aims to simplify a trajectory database $D$ {\chengr to be within a storage budget} such that its simplified database $D'$ preserves the accuracy of query processing {\CHENG for \emph{multiple} types of trajectory queries} as much as possible when compared to the {\chengr that} on $D$. 

\begin{problem}[QDTS]
Given a trajectory database $D$ and a storage budget $W$ indicating a fraction {\chengr $r$} of the original points {\CHENG in $D$} that can be retained, \textbf{Query-Driven Trajectory Simplification} aims to find a trajectory database $D'$ of simplified trajectories, such that the difference between query results on $D$ and $D'$ 
is minimized.
\label{prob:qdts}
\end{problem} 

The QDTS problem relies on 
(1) a query type on a trajectory database and (2) a quality measure 
that captures the difference {\CHENG between the query results on the simplified database and those on the original database}. To establish the former, we review the literature, including the evaluation papers on trajectory simplification~\cite{zhang2018trajectory,cao2006spatio,lin2021error} and a recent trajectory survey~\cite{wang2021survey}, and identify four widely-used queries, namely Range Query, $k$NN Query, Similarity Query, and Clustering. For the latter, we define query-based quality measures by following~\cite{zhang2018trajectory}.

\noindent\textbf{Range Query~\cite{cao2003spatio}.} Given a trajectory database $D$, a range query with {\CHENG parameters} $(q_{x_{min}},q_{x_{max}}, q_{y_{min}}, q_{y_{max}}, q_{t_{min}}, q_{t_{max}})$ finds all trajectories that contain at least one point $p_i=(x_i, y_i, t_i)$ such that $q_{x_{min}} \leq x_i \leq q_{x_{max}}$, $q_{y_{min}} \leq y_i \leq q_{y_{max}}$, and $q_{t_{min}} \leq t_i \leq q_{t_{max}}$.

\smallskip
\noindent\textbf{$k$NN Query~\cite{frentzos2005nearest}.} Given a trajectory database $D$, a $k$NN query takes a {\CHENG query} trajectory $T_q$ and a time window $[t_s,t_e]$ as {\CHENG parameters} and returns a set of $k$ trajectories (denoted by $R$) such that $\forall T_i \in R, \forall T_j \in D-R$,
$\Theta(T_q[t_s,t_e], T_i[t_s,t_e]) \leq \Theta(T_q[t_s,t_e], T_j[t_s,t_e])$,
{\CHENG where $\Theta(\cdot, \cdot)$ represents a dissimilarity measure for trajectories.}
{\CHENG In this paper, } we consider EDR~\cite{chen2005robust} and t2vec~\cite{li2018deep} to instantiate $\Theta(\cdot, \cdot)$, as these represent non-learning and learning based trajectory similarity measures~\cite{wang2021survey,zhang2018trajectory}, respectively. Note that our solution is orthogonal to the dissimilarity measure used.

\smallskip
\noindent\textbf{Similarity Query~\cite{chen2009design}.} Given a trajectory database $D$, a similarity query takes a trajectory $T_q$ and a time window $[t_s,t_e]$ as inputs and returns a set of trajectories (denoted as $R$), such that $d(T_q[i], T_j[i]) \leq \delta$ for any $t_s \leq i \leq t_e$,
where $T_j \in R$, $d(T_q[i], T_j[i])$ is the Euclidean distance between two points $T_q[i]$ and $T_j[i]$ and $\delta$ is a given distance threshold.


\smallskip
\noindent\textbf{Trajectory Clustering~\cite{lee2007trajectory}.} 
Given a trajectory database, {\CHENG trajectory clustering} partitions each trajectory into subtrajectories and then clusters subtrajectories {\CHENG based on some notion of distance among trajectories.} 

\smallskip
\noindent\textbf{Quality Measures.} We use the $F_1$-score for measuring the difference between query results on an original database $D$ and {\CHENG those on} a simplified database $D'$. The idea is to use the results on $D$ as the ground truth and then measure the quality of the results on $D'$ using the $F_1$-score. {\zhengb A larger $F_1$-score indicates a smaller difference between the query results.}

For range, similarity, {\CHENG and $k$NN queries}, we denote by $R_o$ and $R_s$ the trajectory sets returned on $D$ and $D'$, respectively. We define precision (P), recall (R), and $F_1$-score as follows. 
\begin{align}
\text{P} = \frac{|R_o \cap R_s|}{|R_s|} \quad \text{R} = \frac{|R_o \cap R_s|}{|R_o|} \quad F_{1} = 2\cdot\frac{\text{P}\cdot\text{R}}{\text{P}+\text{R}}
\label{eq:f1score}
\end{align}
{\CHENG In particular, for a $k$NN query, the precision, recall, and $F_1$-score are equal since $|R_o| = |R_s| = k$.}


For the {\CHENG trajectory} clustering query, 
{\CHENG we define $R_o$ (resp. $R_s$) to be the set of pairs of trajectories, which are from the same cluster in the results on $D$ (resp. $D'$), and then define the $F_1$-score as above.}

{\CHENG 
\smallskip
\noindent\textbf{QDTS v.s. Existing Trajectory Simplification Problems.}
The QDTS problem differs substantially from the existing error-driven trajectory simplification problems. First, it aims to optimize the data usability (i.e., the query accuracy) directly, as opposed to optimizing an error measure. Second, it targets a database of trajectories and simplifies the trajectories collectively, as opposed to separately. 
}

{\zhengIII{
\smallskip
\noindent\textbf{Remarks.}
We emphasize that we only produce \emph{one} simplified database
{\chengrr and use the simplified database to support \emph{multiple} types of queries including range query, $k$NN query, similarity query, clustering, and possibly others}.
}}
\section{METHODOLOGY}
\label{sec:method}

{\CHENG We propose a new algorithm called \texttt{RL4QDTS} for {\chengb QDTS}.
It starts with the most simplified database, in which each simplified trajectory $T'$ of an original trajectory $T$ consists of only the first and last points of $T$. It then introduces original points into the simplified database iteratively until its budget is exhausted. For better efficiency, it builds an octree on the database of trajectories. Whenever it needs to choose a point, it first chooses a cube in the octree and then chooses a point in that cube. The octree recursively partitions a 2D spatial and 1D temporal space into 8 sub-spaces, which we call (spatial-temporal) cubes. 
{\chengb This is essentially a sequential process of two decision tasks, and therefore, it adopts reinforcement learning (RL) since RL is widely known for its power of handling sequential decision processes.}
Specifically, \texttt{RL4QDTS}
%
%
employs an agent (called Agent-Cube) to traverse the octree to find {\chengr a cube}. 
Then, it employs another agent (called Agent-Point) to choose a point in the cube {\chengr to be inserted} into the simplified database.
%
The decision making processes by the two agents are modeled as Markov decision processes (MDP)~\cite{sutton2018reinforcement} and the MDPs are designed so that the agents cooperatively optimize the query accuracy on the simplified database.
%
%

We present the details of the MDPs of Agent-Cube and Agent-Point in Sections~\ref{sec:mdp-agent-cube} and~\ref{sec:mdp-agent-point}, respectively. 
We then describe how the policies for the two MDPs are learned, in Section~\ref{subsec:policy-learning}.
We finally present the \texttt{RL4QDTS} algorithm that leverages the two agents for simplifying a trajectory database, in Section~\ref{sec:RL4QDTS}.}

{\zheng{\subsection{Agent-Cube: MDP for {\CHENG Choosing} a Cube}
\label{sec:mdp-agent-cube}
Consider the task of choosing a cube. 
{\CHENG Agent-Cube chooses a cube by traversing the octree top-down, starting from the root node. Each time it visits a node, it decides whether to stop.
If it stops, it means that the node's cube is chosen;
otherwise, it 
decides which node among the 8 child nodes to visit. We define the Markov decision process (MDP) of Agent-Cube as follows.}

\smallskip
\noindent
\textbf{(1) States.} 
{\chengr Let $s^c$ denote a state of Agent-Cube's MDP, which we define as follows. We denote by $B_i^j (i>1, 1 \le j \le 8)$ a cube of the octree, which is at the $i^{th}$ level and corresponds to the $j^{th}$ child node of its parent node. We designate $B_1^1$ to denote the cube of the root node.}
{\chengr Consider} that Agent-Cube is {\chengr currently} visiting cube $B_i^j$. For cube $B_i^j$, we use the number of trajectories (denoted by $M_{B_i^j}$) and {\chengr the number of} queries (denoted by $Q_{B_i^j}$) that fall into it, to capture the distributions of the data and queries.
As queries are not available beforehand, 
{\CHENG we synthetically generate a workload of range queries, 
each query location is sampled randomly by following a certain distribution (e.g., data distribution)}. 
Formally, the state $s^{c}$ at a cube $B_i^j$ is defined by its 8 child nodes ($B_{i+1}^1, B_{i+1}^2, ..., B_{i+1}^8$) with two distribution features (data and query) as follows.
\begin{equation}
s^c = \{\frac{M_{B_{i+1}^1}}{M_{B_{i}^j}}, \frac{Q_{B_{i+1}^1}}{Q_{B_{i}^j}}, ..., \frac{M_{B_{i+1}^8}}{M_{B_{i}^j}}, \frac{Q_{B_{i+1}^8}}{Q_{B_{i}^j}}\}, i \ge 1.
\label{eq:state-cube-nor}
\end{equation}
{\CHENG Here, the values of a state are normalized} by dividing by the total numbers of trajectories and queries {\chengr in cube $B_i^j$} (i.e., $M_{B_i^j}$ and $Q_{B_i^j}$) to avoid data scale issues.
{\chengr We explain the intuition of the state design as follows.}
The data values in the states of cubes capture how trajectories are distributed over the cubes. For example, if a cube has only few trajectories and is sparse, an agent tends to select this cube {\CHENG to ensure that} data in that cube is not lost. Similarly, query values in the cubes capture how queries are distributed over the cubes. Intuitively, an agent tends to select a cube with a larger value since {\CHENG the data would serve more queries}.
{\CHENG We note that in cases we have some knowledge of the query workload for testing (e.g., its distribution), we can generate query workloads by following the distribution {\chengr for training}; in cases we have no knowledge of the query workload for testing, we can generate a query workload by following the data distribution for training. In our experiments, we conduct experiments {\chengr which verify to some extent} the transferability of our method for cases where the query workload for testing does not follow that of the one used for training.}

\smallskip
\noindent
\textbf{(2) Actions.}
Let $a^c$ denote an action of Agent-Cube's MDP. With the currently visited cube being $B_i^j$, we define two possible types of action: (1) Proceed to visit one of the 8 child nodes, and (2) Stop the traversal and return the current cube to Agent-Point, to choose a point within the chosen cube.
{\CHENG Formally, } $a^c$ is defined as follows.
\begin{equation}
    a^c = k~~(1 \le k \le 9).
\end{equation}
{\CHENG Here, $a^c = 1,2,...,8$ means to traverse one of the 8 child nodes of the current one and $a^c = 9$ means to stop at the current node.}
{\CHENG Furthermore, we constrain the action space by only considering the cubes that involve trajectories.} Suppose we take an action $a^c=k$, which corresponds {\chengr to} one of the two transition cases. Case 1: it explores the next cube $B_{i+1}^k$ if $1 \le k \le 8$, and a new state at cube $B_{i+1}^k$ can be computed using Equation~\ref{eq:state-cube-nor}. Case 2: it stops and returns to Agent-Point if $k=9$. More details {\CHENG of Agent-Point} are presented in Section~\ref{sec:mdp-agent-point}.

\smallskip
\noindent
\textbf{(3) Rewards.} When the action is to explore one of the 8 child nodes, the reward cannot be immediately observed, since no point has been inserted into the simplified database. {\CHENG When the action is to choose the current cube for Agent-Point to choose a point within the cube, the simplified database would be updated and some reward signal can be acquired (e.g., by measuring the {\chengr difference} between the query accuracy on the original database and that on the {\chengr updated} simplified {\chengr database}).}
{\CHENG In summary, Agent-Cube would finally choose a cube for Agent-Point and then acquire a certain reward signal. Therefore, } 
we make Agent-Cube and Agent-Point share the same rewards, since they cooperate {\chengr towards} the same objective, i.e., learning a query-aware policy such that a simplified database preserves the query accuracy as much as possible compared to the original database.
In particular, we set the reward {\CHENG of an action by Agent-Cube} to be equal to that of the following action of choosing a point within the selected cube {\chengrr by Agent-Point}. More details {\CHENG of the reward definition {\chengrr of Agent-Point}} are presented in Section~\ref{sec:mdp-agent-point}.}}
\subsection{Agent-Point: MDP for Choosing a Point}
\label{sec:mdp-agent-point}
We {\CHENG denote} the {\CHENG chosen} cube by Agent-Cube as $B$ for simplicity. 
{\CHENG Next, we define the MDP of Agent-Point for choosing a point within $B$ to introduce to the database.}
%

\smallskip
\noindent
\textbf{(1) States.} 
{\chengr Let $s^p$ denote a state of Agent-Point's MDP, which we define as follows.} Let $N_B$ (resp. $M_B$) denote the number of points (resp. trajectories) in the cube $B$. 
To define the state, one idea is to {\CHENG incorporate} all $N_B$ points.
However, this idea has two issues. (1) The definition in this way is $N_B$-dependent, which is not suitable for other cases when the number of points is not $N_B$. 
(2) $N_B$ is generally very large. With this definition, the state space would be huge and the model is hard to train.
%


%
We design the states such that these two issues are avoided as follows. 
First, 
let $p_{s_a}^{T_i}$ and $p_{s_b}^{T_i}$ denote the first point and last point of {\CHENG a trajectory} ${T_i}$ within the cube, respectively.
%
For each point $p_{s_j}^{T_i}$ in the cube {\CHENG with $s_a \leq s_j \leq s_b$}, we define a pair of two values, denoted by $v(p_{s_j}^{T_i})$, as follows. {\chengrr
\begin{equation}
v(p_{s_j}^{T_i}) = ( v_s( p_{s_j}^{T_i}), v_t( p_{s_j}^{T_i})).
\label{eq:state-point}
\end{equation}}
The first value, denoted by $v_s(p_{s_j}^{T_i})$, is equal to the ``spatial'' distance between $p_{s_j}^{T_i}$ and the synchronous point {\CR (i.e., the location at the time of $p_{s_j}^{T_i}$ based on {\zhengII{the segment linking the points immediately before and after $p_{s_j}^{T_i}$ in the trajectory $T_i$.}})}
The second value, denoted by $v_t(p_{s_j}^{T_i})$, is equal to the ``temporal'' difference between the time of $p_{s_j}^{T_i}$ and the time of $p_{s_j}^{T_i}$'s closest 
point on {\zhengII{the segment linking the points immediately before and after $p_{s_j}^{T_i}$ in the trajectory $T_i$.
}}
%
%
The intuition of the two values is to capture the features of the point $p_{s_j}^{T_i}$ from both the spatial and temporal aspects given the context of trajectory simplification.

Among all points in each trajectory $T_i$, we then find a point (denoted as $p_{s_*}^{T_i}$) which has the maximum $v_s$, where $s_*$ denotes its index. That is,
\begin{equation}
    s_* = \argmax_{s_a \leq s_j \leq s_b} v_s(p_{s_j}^{T_i}).
\label{eq:state-traj}
\end{equation}

Finally, the state $s^p$ of Agent-Point is defined as the set of $K$ largest $v_s$ values of $v(p_{s_*}^{T_i})$ among the $B_M$ trajectories, that is
\begin{equation}
s^p = \{v(p_{s_*}^{T_{\pi(1)}}), v(p_{s_*}^{T_{\pi(2)}}), ..., v(p_{s_*}^{T_{\pi(K)}})\},
\label{eq:state}
\end{equation}
where $\pi$ denotes the permutation of $T_1, T_2, ..., T_{M_B}$ such that 
{\CHENG $v_s(p_{s_*}^{T_{\pi(1)}})$, $ v_s(p_{s_*}^{T_{\pi(2)}})$, ..., $v_s(p_{s_*}^{T_{\pi(M_B)}})$ is sorted {\chengrr in a descending order}.}
$K$ ($K\le M_B$) is a hyper-parameter that can be tuned empirically to control the size of the state space. {\zhengI{Note that if a point has been introduced in the database, the point will not be used for the state definition.}} {\CR We also consider the state based on the set of $K$ largest $v_t$ values, and they perform worse than based on $v_s$ empirically. 
}

{\zhengI{Here, we refer to an example for illustrating the state definition. Consider a cube $B_3^{4}$ (the bottom right node at the third tree level) in Figure~\ref{fig:example_qdts}. It contains two points $p_5$ and $p_8$ for the definition. For $p_5$ (resp. $p_8$), we calculate the values as $(1.6, 0.5)$ (resp. $ (1.3, 0.7)$) for capturing the spatial and temporal distances with respect to its simplified segment $\overline{p_4p_6}$ on $T_2$ (resp. $\overline{p_7p_9}$ on $T_3$). Then, the state is constructed as $s^p=\{(1.6, 0.5), (1.3, 0.7)\}$ with the setting of $K=2$.
}}

Our state design avoids the two aforementioned issues, where $K$ is generally much smaller than $N_B$ or $M_B$. With this design, a state has a fixed size that is independent from the number of trajectories in the cube.



\smallskip
\noindent
\textbf{(2) Actions.}
Let $a^p$ denote an action of Agent-Point.
The design of actions is consistent with the design of state $s^p$. Specifically, the actions are defined as follows:
\begin{equation}
    a^p = k~~(1 \le k \le K),
\end{equation}
where action $a^p = k$ means to insert point $p_{s_*}^{T_{\pi(k)}}$ into $D'$.

\smallskip
\noindent
\textbf{(3) Rewards.} 
Since our objective is to obtain a simplified database that serves queries more effectively (i.e., minimizing the difference between the query results on the original database and those on the simplified database),  the reward is expected to reflect the improvement of query performance as more points are included in the simplified database.
%
%
{\chengr To this end}, we {\chengr use the} query workloads {\chengr that have been used for defining the states} (e.g., {\chengr we use} a set of range queries, where each query location is randomly sampled by following the data distribution). 
One option is to perform the queries after each point is inserted to the simplified database, which is associated with the transition from the current state $s^p$ to the next state $s^{p'}$ when an action $a^p$ is taken. However, it would be prohibitively costly to perform queries for each inserted point.
In addition, since the simplified database $D'$ has not been fully constructed, the query improvement with inserting just one point is often negligible and it is hard to demonstrate the quality of the action. 

In our design, we choose to perform the queries after $\Delta$ (e.g., $\Delta=50$) points are inserted for achieving accumulative effects. Specifically, we denote the reward by $R$. At state $s_i^p$, we consider the simplified database 
({\CHENG denoted by} $D'$). 
At state $s_{i+\Delta}^p$, we {\CHENG consider} the simplified database {\chengrr again}
({\CHENG denoted by} $D''$). We then define the reward $R$ as follows.
\begin{equation}
    R = \mathit{diff}(Q(D), Q(D')) - \mathit{diff}(Q(D), Q(D'')),
\end{equation}
where $\mathit{diff}(Q(D), Q(D'))$ measures the difference between the queries on the original database $D$ and the simplified database $D'$. The intuition is that if the difference for the simplified database $D''$ is smaller, then the reward is larger. 
Furthermore, {\CHENG we make} the reward $R$ be shared by all transitions that {are involved when traversing} from $s_i^p$ to $s_{i+\Delta}^p$ {\CHENG as well as those of Agent-Cube that are involved in this process.}

With the above reward definition, 
the objective of the MDP, i.e., maximizing the accumulative rewards, would be equivalent to that of the QDTS problem, i.e., minimizing the difference between queries on the original database and {\CHENG those on} the simplified database. To see this, suppose we traverse a sequence of $N'$ states $s_1^p, s_2^p, ..., s_{N'}^p$ {\chengr (for simplicity, we assume $\Delta = 1$ for this analysis)}. Correspondingly, we receive a sequence of rewards $R_1, R_2, ..., R_{N'-1}$. We assume that the future rewards are accumulated without discounted rates, and thus the accumulative reward is calculated as follows.
\begin{equation}
\begin{aligned}
\label{reward}
\sum_{t=1}^{N'-1}R_t &= \sum_{t=1}^{N'-1}{\chengrr (}\mathit{diff}(Q(D), Q(D'_{t})) - \mathit{diff}(Q(D), Q(D''_{t})){\chengrr )} \\
&= \mathit{diff}(Q(D), Q(D'_{1})) - \mathit{diff}(Q(D), Q(D''_{N'-1})) \\
&= C - \mathit{diff}(Q(D), Q(D''_{N'-1})),
\end{aligned}
\end{equation}
where $D'_{t}$ (resp. $D''_{t}$) denotes the simplified database at the state $s_t^p$ before (resp. after) the action $a_t^p$ is performed.
We regard the initial term $\mathit{diff}(Q(D), Q(D'_{1}))$ as a constant $C$ and no points have been inserted at that state.
Therefore, the objective of the MDP is to maximize $C-\mathit{diff}(Q(D), Q(D''_{N'-1}))$ or equivalently to minimize $\mathit{diff}(Q(D),$ $Q(D''_{N'-1}))$, which is exactly the objective of QDTS.

\subsection{Policy Learning via DQN}
\label{subsec:policy-learning}
The core problem of a MDP is to find an optimal policy, which guides an agent to choose an action at a specific state, such that the accumulative reward is maximized. 
{\CHENG Considering that the states in our MDPs are continuous,}
we adopt the Deep-Q-Networks (DQN)~\cite{mnih2013playing} for learning a policy from the MDPs of Agent-Cube and Agent-Point. Specifically, we adopt the deep Q learning with replay memory~\cite{mnih2013playing} for learning the policy, denoted by $\pi_{\theta^{c}}(a|s^c)$ for Agent-Cube (resp. $\pi_{\theta^{p}}(a|s^p)$ for Agent-Point).
{\chengrr The policy} samples an action $a$ at a given state $s^c$ (resp. $s^p$) via DQN, whose parameters are denoted by $\theta^{c}$ (resp. $\theta^{p}$). {\zhengII{We note that other RL algorithms such as policy gradient can also be used for continuous state MDPs. 
}}

{\zheng{

{\zheng{
\begin{algorithm}[t]
	\caption{The framework of \texttt{RL4QDTS}}
	\label{alg:rl4qdts_model}
		\DontPrintSemicolon
        \SetKwFunction{FMain}{RL4QDTS}
        \SetKwProg{Pn}{Function}{:}{\KwRet{
        $D'$}}
        \Pn{\FMain{$D=\langle T_1, T_2, ..., T_M \rangle$, $W$}}{
		Build an octree $OT$ for $D$;\\
		\For{i=1,2,...,M}{
		Insert point $p_1^{T_i}$ and $p_{|T_i|}^{T_i}$ into $D'$; \\
		}
        \For{i=2M+1,2M+2,...,W}{
        $B\leftarrow\texttt{Agent-Cube}(OT)$; \\ 
        $D'\leftarrow\texttt{Agent-Point}(B, D')$; \\ 
		}}
\end{algorithm}

\begin{algorithm}[t]
	\caption{The \texttt{Agent-Cube}}
	\label{alg:rl4qdts_cube}
    \DontPrintSemicolon
    \SetKwFunction{FMain}{Agent-Cube}
    \SetKwProg{Pn}{Function}{:}{\KwRet{$B$}}
    \Pn{\FMain{$OT$}}{
        $h \leftarrow 1$ and $l \leftarrow 1$;\\
        \While{true}{
        Construct a state $s_h^c \leftarrow \{\frac{M_{B_{h+1}^1}}{M_{B_{h}^l}}, \frac{Q_{B_{h+1}^1}}{Q_{B_{h}^l}}, ..., \frac{M_{B_{h+1}^8}}{M_{B_{h}^l}}, \frac{Q_{B_{h+1}^8}}{Q_{B_{h}^l}}\}$;\\
        Sample an action $a_h^c \sim \pi_{\theta^c}(a|s_h^c)$;\\
         \uIf{$a_h^c=9$ or {\CR $B_h^l$ is a cube of a leaf node}}{
         $B \leftarrow B_h^l$;\\
         \textbf{Break;}\\
         }
         \Else{
         $h \leftarrow h+1$ and $l \leftarrow a_h^c$;\\
         {\chengr \textbf{Continue}};\\
         }
        }
    }
\end{algorithm}

\begin{algorithm}[t]
	\caption{The \texttt{Agent-Point}}
	\label{alg:rl4qdts_point}
    \DontPrintSemicolon
    \SetKwFunction{FMain}{Agent-Point}
    \SetKwProg{Pn}{Function}{:}{\KwRet{$D'$}}
    \Pn{\FMain{$B, D'$}}{
		Compute $v(p_{s_*}^{T_j})$ ($1 \le j \le M_{B}$) for $T_j \in B$;\\
        {\CHENG Maintains a descending permutation $\pi$ of the values with a max-priority queue;}\\
		Construct a state $s^p \leftarrow \{v(p_{s_*}^{T_{\pi(1)}}), v(p_{s_*}^{T_{\pi(2)}}), ..., v(p_{s_*}^{T_{\pi(K)}})\}$;\\
		Sample an action $a^p \sim \pi_{\theta^p}(a|s^p)$;\\
		Insert the point $p_{s_*}^{T_{\pi(k)}}$ into $D'$ where $a^p=k$ ($1 \le k \le K$);\\
	}
\end{algorithm}


\begin{figure*}[]
	\centering
	\begin{tabular}{cc}
		\centering
		\hspace{-3mm}
		\begin{minipage}{0.53\textwidth}
			\centering
			\includegraphics[width=\textwidth]{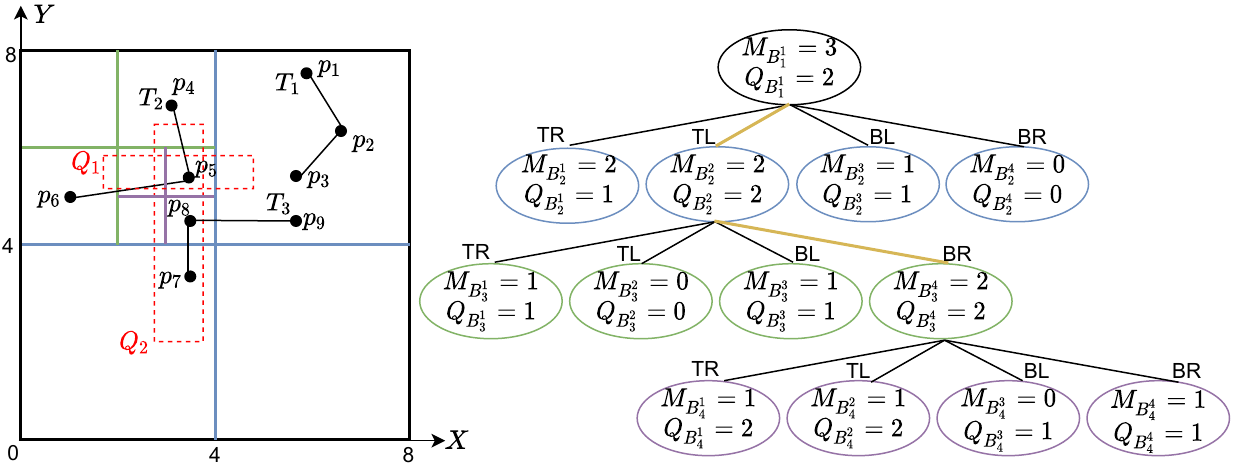}
			\label{fig:input}
		\end{minipage}\hspace{-3mm}
		&
		\hspace{-1mm}
		\begin{minipage}{0.4\textwidth}
			\setlength{\tabcolsep}{2.2pt}
			\centering
\begin{tabular}{|c|ccc|}
\hline
Initial & \multicolumn{3}{c|}{Insert $p_1,p_3,p_4,p_6,p_7,p_9$ into $D'$}                                                                                                                                                                \\ \hline
Cube    & \multicolumn{1}{c|}{Ag.}                 & \multicolumn{1}{c|}{State}                                                                                                                        & Action                          \\ \hline
$B_1^1$ & \multicolumn{1}{c|}{C}                   & \multicolumn{1}{c|}{$\{\frac{2}{3}, \frac{1}{2}, \frac{2}{3}, \frac{2}{2}, \frac{1}{3}, \frac{1}{2}, \frac{0}{3}, \frac{0}{2}\}$}                 & Explore $B_2^2$                 \\ \hline
$B_2^2$ & \multicolumn{1}{c|}{C}                   & \multicolumn{1}{c|}{$\{\frac{1}{2}, \frac{1}{2}, \frac{0}{2}, \frac{0}{2}, \frac{1}{2}, \frac{1}{2}, \frac{2}{2}, \frac{2}{2}\}$}                 & Explore $B_3^4$                 \\ \hline
$B_3^4$ & \multicolumn{1}{c|}{C}                   & \multicolumn{1}{c|}{$\{\frac{1}{2}, \frac{2}{2}, \frac{1}{2}, \frac{2}{2}, \frac{0}{2}, \frac{1}{2}, \frac{1}{2}, \frac{1}{2}\}$}                 & Sample $B_3^4$                  \\ \hline
$B_3^4$ & \multicolumn{1}{c|}{P}                   & \multicolumn{1}{c|}{$\{(1.6, 0.5), (1.3, 0.7)\}$}                                                                                                 & Choose $p_5$                    \\ \hline
Output  & \multicolumn{3}{c|}{\begin{tabular}[c]{@{}c@{}}Return $D'= \langle T'_1, T'_2, T'_3 \rangle$, where \\ $T_1'= \langle p_1, p_3 \rangle$, $T_2'= \langle p_4, p_5, p_6 \rangle$, $T_3'= \langle p_7, p_9 \rangle$\end{tabular}} \\ \hline
\end{tabular}
        \end{minipage}
	\end{tabular}
 	\vspace*{-4mm}
    \caption{{\zhengb A running example of \texttt{RL4QDTS}, where we use a quadtree to simplify demonstration by eliminating the temporal dimension. \textbf{Left}: Three trajectories $T_1$, $T_2$, and $T_3$ exist in the original database $D$, along with two range queries $Q_1$ and $Q_2$ from the query workload. \textbf{Middle}: Quadtree nodes are labeled as TR, TL, BL, and BR (Top Right, Top Left, Bottom Left, and Bottom Right) with different colors representing quadtree levels. Agent-Cube {\CHENG traverses} through nodes following yellow lines. \textbf{Right}: MDPs of Agent-Cube (C) and Agent-Point (P) for inserting point $p_5$ into the simplified database $D'$.}}
	\label{fig:example_qdts}
    \vspace*{-5mm}
\end{figure*}

\subsection{The \texttt{RL4QDTS} Algorithm}
\label{sec:RL4QDTS}
Algorithm~\ref{alg:rl4qdts_model} details the framework of \texttt{RL4QDTS} with the {\CHENG learned policies of Agent-Cube and Agent-Point} for the QDTS problem. 
Specifically, \texttt{RL4QDTS} starts by building an octree for the original trajectory database $D$ (line 2), and then inserts the first and the last points of each trajectory into a simplified trajectory database $D'$ (lines 3 -- 5).
The remaining budget $W-2M$ is utilized in lines 6 -- 9. First, it calls Agent-Cube (to be presented in Algorithm~\ref{alg:rl4qdts_cube}) to {\CHENG choose} a cube $B$, and then the cube is fed into Agent-Point (to be presented in Algorithm~\ref{alg:rl4qdts_point}) for updating $D'$. The process continues until the budget is exhausted. The \texttt{RL4QDTS} algorithm returns $D'$, which contains $W$ points (line 10).

Agent-Cube in Algorithm~\ref{alg:rl4qdts_cube} first initializes the indexes $h$ and $l$ to 
{\chengr indicate a cube $B_h^l$}
(line 2). 
To sample a cube, in lines 3 -- 13,
it constructs a state $s_h^c$ using Equation~\ref{eq:state-cube-nor} (line 4), and samples an action $a_h^c$ with the learned policy $\pi_{\theta^c}(a|s_h^c)$, which takes $s_h^c$ as input (line 5). If the action is $a_h^c=9$ {\CR or the exploration reaches the last level of the tree}, it breaks and returns the current cube denoted by $B$ (lines 6 -- 8); otherwise, it updates the indexes by $h \leftarrow h+1$ and $l \leftarrow a_h^c$, and explores the next cube $B_h^l$ (lines 9 -- 12). 

Agent-Point in Algorithm~\ref{alg:rl4qdts_point} takes a cube $B$ as input and computes the value of $v(p_{s_*}^{T_j})$ for each trajectory $T_j$ by Equations~\ref{eq:state-point} and~\ref{eq:state-traj}, where $1\le j \le M_{B}$ (line 2). Then, {\CHENG it maintains a descending permutation $\pi$ of the values with a max-priority queue (line 3).}
Next, it constructs a state $s^p$ by Equation~\ref{eq:state} (line 4), and samples an action with the learned policy $\pi_{\theta^p}(a|s^p)$, which takes $s^p$ as input (line 5). Let $a^p=k$ ($1 \le k \le K$) denote the sampled action. It then takes the action by inserting the point $p_{s_*}^{T_{\pi(k)}}$ into $D'$ (line 6). 

We illustrate the \texttt{RL4QDTS} Algorithm with the running example in Figure~\ref{fig:example_qdts}. Here, we use a quadtree (instead of an octree) by ignoring the temporal dimension of trajectory points for ease of demonstration. The input is a trajectory database $D = \langle T_1, T_2, T_3 \rangle$ with storage budget $W=7$. Suppose the range query workload {\chengr involves} $Q_1$ and $Q_2$, where {\chengr we have} $Q_1(D)=\langle T_2\rangle$ and $Q_2(D)=\langle T_2, T_3\rangle$ based on the original database $D$.
We build a quadtree and record the number of trajectories ($M_B$) and queries ($Q_B$) in the tree nodes. (1) The algorithm first inserts the first and the last points of each trajectory into $D'$, meaning that the remaining budget is one point (i.e., 7-2*3=1). (2) Then, Agent-Cube starts at the root node $B_1^1$ and constructs its state by observing the four child nodes. It takes the action to explore node $B_2^2$ (Top Left in the figure). (3) Similarly, at $B_2^2$, it takes the action to explore node $B_3^4$ (Bottom Right). (4) At $B_3^4$, Agent-Cube receives the action of providing $B_3^4$ to Agent-Point. (5) Agent-Point constructs the state at cube $B_3^4$ and takes the action to insert point $p_5$ into $D'$. (6) Finally, the algorithm breaks from the loop and returns the simplified database $D'$ since the budget is exhausted. We observe that $D'$ outputs the same results (i.e., $Q_1(D')=\langle T_2\rangle$ and $Q_2(D')=\langle T_2, T_3\rangle$) as when querying $D$, indicating the preservation of query accuracy.

In addition, we {\chengr develop} two techniques to {\CHENG enhance} the effectiveness and efficiency of $\texttt{RL4QDTS}$. First, we constrain the octree traversal of Agent-Cube by a maximum tree depth $E$. If Agent-Cube reaches this level, it returns the currently visited cube to Agent-Point.
The rationale is to prevent a very long {\CHENG traversal path} for Agent-Cube, {\CHENG since in this case}, it is difficult to train a policy to converge - recall that the reward is {\chengr computed with delays}. The benefits are verified in experiments. 
Second, we set a start level $S$ {\CHENG so that} the Agent-Cube {\CHENG starts traversing the octree by} randomly sampling a cube following the query distribution ({\zhengII{the one that has been used for defining states}}) from the start level $S$. The number of points in a cube decreases as the tree level increases. If Agent-Cube stops at the root level, Agent-Point will operate on all points in the database. Hyperparameter $S$ can be used to avoid returning cells with excessive numbers of points. 

\smallskip
\noindent\textbf{Time complexity.}
The time complexity of the \texttt{RL4QDTS} algorithm is $O(N+W(n+\log M_B))$, where $N$, $W$, $n$, and $M_B$ denote the total number of points in the original database, the storage budget, the maximum number of points in the input trajectories, and the maximum number of trajectories in the data cubes. 
Specifically, it takes $O(N)$ time to build an octree on the original database with maximum tree depth $E$, which is a small constant~\cite{samet1984quadtree}.
The {\chengr part of} processing of the remaining $W-2M$ points dominates the complexity, including (1) {\CHENG choosing} a cube by Agent-Cube with cost $O(1)$, which explores the octree 
{\CHENG for a bounded number of levels;}
(2) computing the values {\chengr by} Agent-Point with cost $O(n)$; (3) maintaining the min-priority queue with cost $O(\log M_B)$; (4) constructing a state, sampling an action, and inserting a point by Agent-Point with cost $O(1)$ assuming $K$ is a small constant.
We note that the \texttt{RL4QDTS} algorithm has the same complexity as the error-driven algorithms~\cite{hershberger1992speeding,marteau2009speeding,wang2021trajectory} for simplifying a set of trajectories. In addition, we note that simplification is normally performed once offline, after which the simplified database is used for online querying.
}}

{\chengrr
\smallskip
\noindent\textbf{Remarks.}
We train \texttt{RL4QDTS} with range queries only and then test it for different types of queries (including range query, kNN Query, Similarity Query, and Clustering) without retraining the model. The rationale is that the range query is a simple yet basic one and by training the model with range queries, the model would learn to capture essential spatial and temporal patterns of trajectories when simplifying them, which would then be useful for other types of queries. We follow this strategy and verify the transferability of our model among different types of queries in experiments.
}}}
\section{EXPERIMENTS}
\label{sec:experiment}


\subsection{Experimental Setup}
\label{sec:setup}
\textbf{Dataset.} We conduct the experiments on {\chengr four} real-world trajectory datasets, Geolife~\footnote{\url{https://www.microsoft.com/en-us/research/publication/geolife-gps-trajectory-dataset-user-guide/}}, T-Drive~\footnote{\url{https://www.microsoft.com/en-us/research/publication/t-drive-trajectory-data-sample/}}, Chengdu~\footnote{\url{https://drive.google.com/file/d/1onzDFpbD9OOfvOK7jHJ6Tpi2V4oKfxXR/view?usp=sharing}} and OSM~\footnote{\url{https://star.cs.ucr.edu/?OSM/GPS\#center=43.6,-56.1\&zoom=2}}.
Geolife contains trajectories from 182 users during a period of five years (2007 -- 2012). 
T-Drive contains trajectories from 10,357 taxis in Beijing over a period of one week. 
{\zheng{Chengdu contains taxi trajectories {\zhengII{from 2016-11-01 to 2016-11-07}}, released by DiDi Chuxing.}}
{\zhengII{OSM is used to test the scalability, which contains three billion points, released by the community on OpenStreetMap.}}
The datasets are widely used in previous trajectory simplification studies~\cite{zhang2018trajectory,wang2021trajectory,long2014trajectory}, and detailed statistics are shown in Table~\ref{tab:dataset}.

\begin{table}[t]
\caption {Dataset statistics.} \label{tab:dataset}
\vspace{-3mm}
\setlength{\tabcolsep}{3pt}
\begin{tabular}{c|c|c|c|c}
\hline
\textbf{Statistics}                & \textbf{Geolife} & \textbf{T-Drive} & \textbf{Chengdu} & \textbf{OSM} \\ \hline
 \# of trajectories                 & 17,621           & 10,359   & 179,756 &513,380 \\
Total \# of points                 & 24,876,978       & 17,740,902 & 32,151,865 &2,913,478,785 \\
Ave. \# of pts per traj & 1,412            & 1,713 & 178 &5,675 \\
Sampling rate                      & 1s $\sim$ 5s          & 177s  & 2s $\sim$ 4s &53.5s \\
Average length                   & 9.96m            & 623m & 25m &180m \\ \hline
\end{tabular}
\vspace*{-7mm}
\end{table}

\begin{figure*}
	\hspace{-3mm}
	\centering
	\begin{tabular}{c c c}
	  \begin{minipage}{4cm}
        \includegraphics[width=14cm]{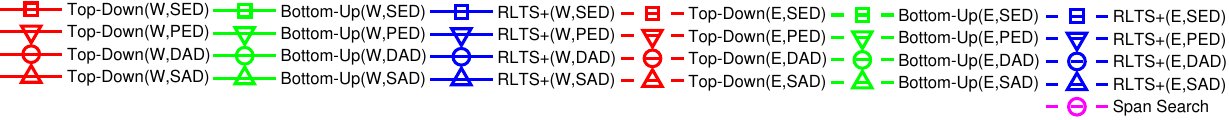}
      \end{minipage}
      &
      \\
		\begin{minipage}{0.3\linewidth}
			\includegraphics[width=\textwidth]{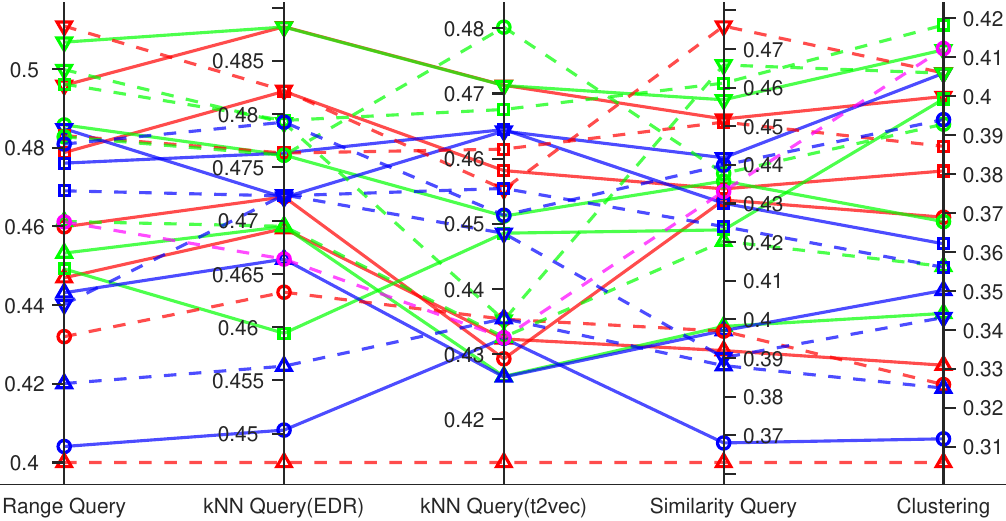}
		\end{minipage}\hspace{-1mm}
		&
		\begin{minipage}{0.3\linewidth}
			\includegraphics[width=\textwidth]{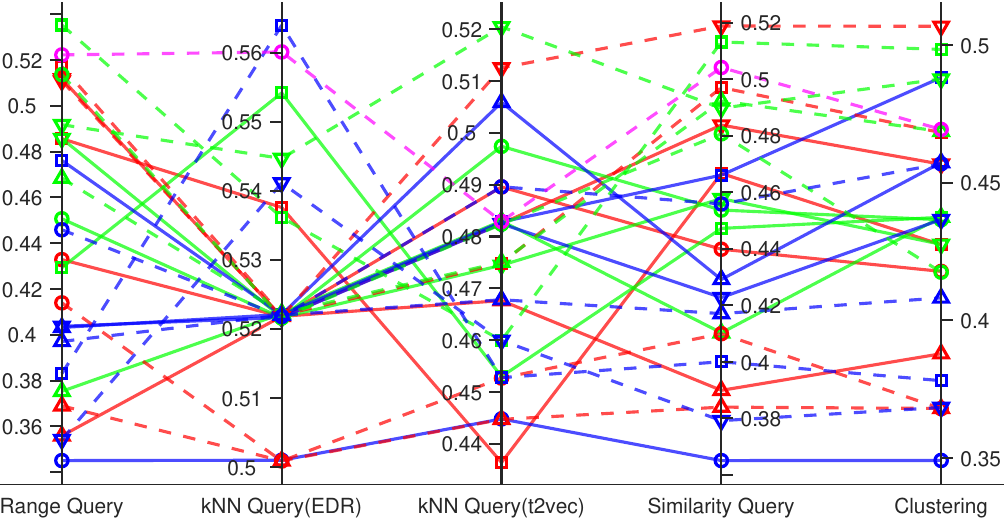}
		\end{minipage}\hspace{-1mm}
		&
		\begin{minipage}{0.3\linewidth}
			\includegraphics[width=\textwidth]{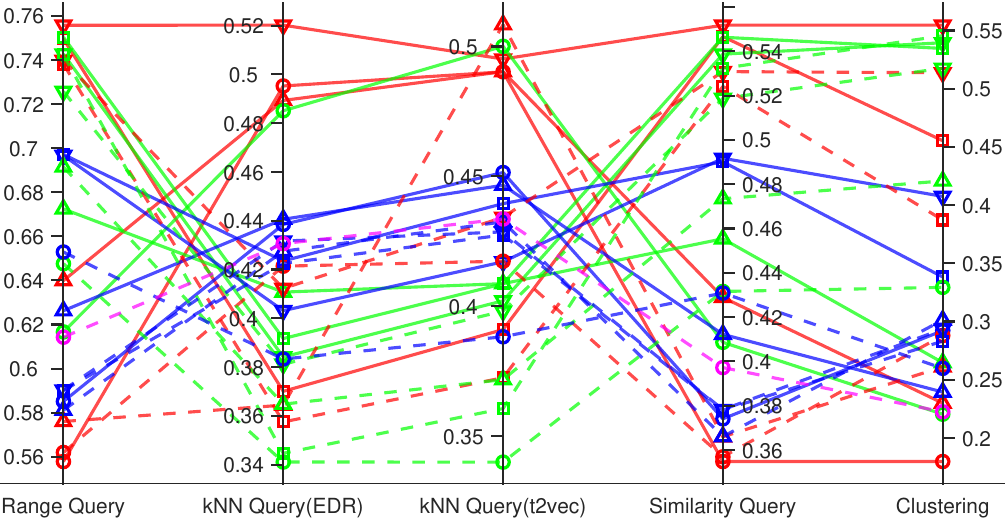}
		\end{minipage}\hspace{-1mm}
		\\
		(a) Data distribution
		&
		(b) Gaussian distribution
		&
		(c) Real distribution
	\end{tabular}
    \vspace{-2mm}
	\caption{{\zhengII{Skyline selection with existing algorithms.}}}\label{fig:skyline}
	\vspace{-7mm}
\end{figure*}

\smallskip
\noindent\textbf{Baselines.} 
{\CHENG In the literature, no algorithms have been proposed for the QDTS problem.
Given that the EDTS problem takes a storage budget for a trajectory as input, we consider existing algorithms that have been proposed for EDTS as baselines in our experiments.} Specifically, we consider four {\CHENG algorithms, namely} Top-Down~\cite{hershberger1992speeding}, Bottom-Up~\cite{marteau2009speeding}, RLTS+~\cite{wang2021trajectory}, and Span-Search~\cite{long2014trajectory}. Among them, Top-Down, Bottom-Up, and RLTS+ are general frameworks that can be applied with different error measures, while Span-Search works with DAD only. We adapt Top-Down, Bottom-Up, and RLTS+ in two ways. The first is to simplify \underline{e}ach trajectory in the database one by one by calling one of the algorithms (this adaptation is denoted as ``E''). The second is to consider the database as a \underline{w}hole and simplify the database by inserting or dropping points among all points in the database as it simplifies a trajectory (this adaptation is denoted as ``W''). In summary, for each of the algorithms Top-Down, Bottom-Up, and RLTS+, we obtain 8 (= $4\cdot 2$) adaptations as baselines, each corresponding to a combination of an error measure SED, PED, DAD, or SAD, and an adaptation method (``E'' and ``W''). In total, we have 25 baselines including 24 (= $3\cdot 8$) adaptations of Top-Down, Bottom-Up, and RLTS+ and 1 adaption of Span-Search (we note that for Span-Search, the ``W'' adaptation is not possible). 

\smallskip
\noindent\textbf{Evaluation Platform.}
We implement \texttt{RL4QDTS} and the baselines in Python 3.6. Experiments are conducted on a 10-cores server with an Intel(R) Core(TM) i9-9820X CPU @3.30GHz 64.0GB RAM and an Nvidia GeForce RTX 2080 GPU.
The datasets and code are available via the link\footnote{\url{https://github.com/zhengwang125/Query-TS}}. 

\smallskip
\noindent \textbf{Model Training and Parameter Settings.} 
Agent-Cube is implemented using a two-layered feedforward neural network (FNN) with 25 neurons in the first layer using the tanh activation function, and 9 neurons in the second layer corresponding to the action space with a linear activation function. The hyperparameters $S$ and $E$ are set to 9 and 12, respectively, based on empirical findings. For Agent-Point, a two-layered FNN is used, with 25 neurons in the first layer using the tanh activation function and $K$ neurons in the second layer corresponding to the action space with a linear activation function, where $K$ is set to 2. Batch normalization is employed in the neural networks to avoid data scale issues.

For training, we randomly sample 6,000 trajectories from Geolife (resp. 6,000 trajectories from T-Drive, 48,000 trajectories from Chengdu, and 6,000 trajectories from OSM), with the remaining trajectories used for testing. From these, 12 databases are randomly prepared, each containing 500 (resp. 500, 4,000, and 500) trajectories. Five episodes are generated for each database to train the policy, and the best model is chosen during training. Around one million transitions are produced in the training process, which is sufficient for training a good policy with reasonable time based on empirical findings. Additionally, we set $\Delta = 50$, i.e., for every 50 points that have been inserted, we perform 100 range queries, each with a spatial region of 2km by 2km and a temporal duration of 7 days to construct states and acquire rewards. Range queries are solely used for training as it involves both spatial and temporal dimensions. We vary the query distributions across three types: (1) the data distribution, (2) the Gaussian distribution (with parameters $\mu=0.5$ and $\sigma=0.25$), and (3) the real distribution for the Chengdu dataset, which generates queries near pickup and dropoff locations similar to real queries in ride-hailing services. The discount rate is set to 0.99, and the \texttt{RL4QDTS} model is trained using Adam stochastic gradient descent with an initial learning rate of 0.01. For $\epsilon$-greedy in DQN, the minimal $\epsilon$ is set to 0.1 with a decay of 0.99, and the replay memory size is set to 2000.

For testing, \texttt{RL4QDTS} involves some random cube sampling at the start level. We run the algorithm 50 times and collect averages and standard deviations of query metrics for each result. Range queries are set as cubes with a spatial region of 2km by 2km and a temporal duration of 7 days. $k$NN queries have a window length of 7 days with $k=3$ and use EDR and t2vec as similarity measures (EDR threshold: 2km, t2vec settings as described in~\cite{li2018deep}). Similarity queries have a distance threshold of 5km. For clustering, we adopt the TRACLUS algorithm as described in the original paper~\cite{lee2007trajectory}. {\CR We are not using GPU for the implementation.}

%


\begin{figure*}[ht]
	\centering
	\hspace*{-4mm}
	\begin{tabular}{c c c c c}
		\hspace{-3mm}
        \begin{minipage}{0.2\linewidth}
			\includegraphics[width=\linewidth]{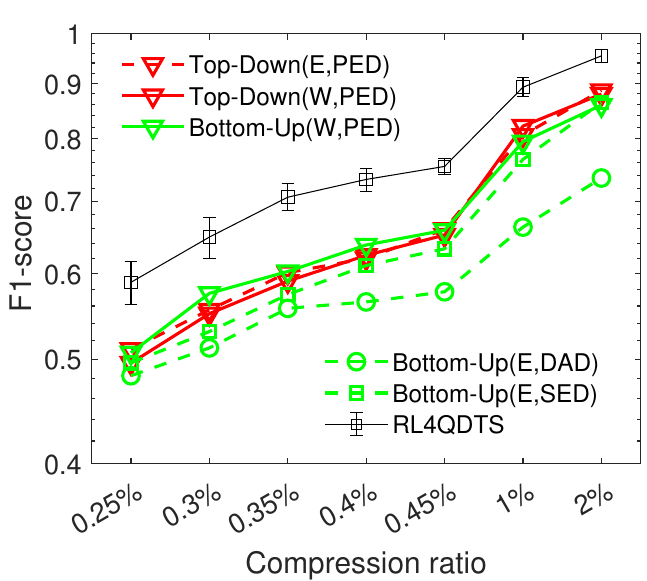}
		\end{minipage}\hspace{-3mm}
		&
		\begin{minipage}{0.2\linewidth}
			\includegraphics[width=\linewidth]{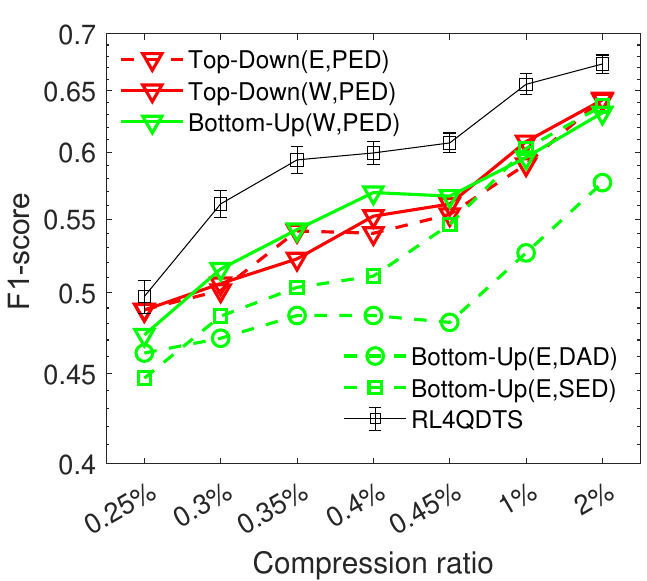}
		\end{minipage}\hspace{-3mm}
		&
		\begin{minipage}{0.2\linewidth}
			\includegraphics[width=\linewidth]{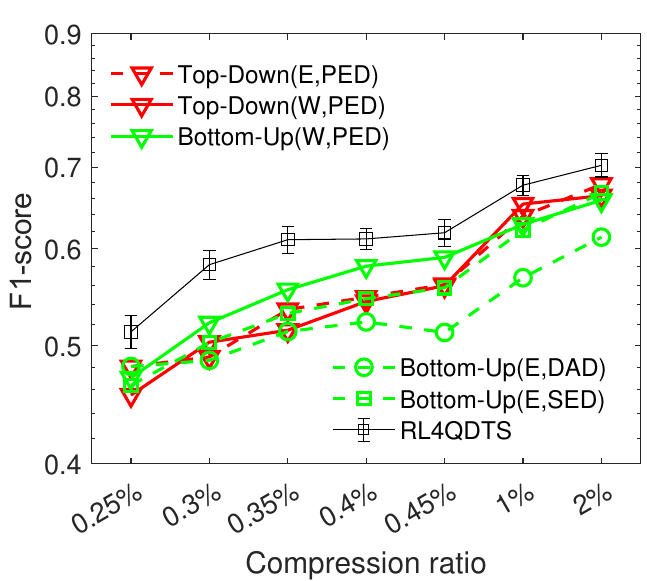}
		\end{minipage}\hspace{-3mm}
		&
		\begin{minipage}{0.2\linewidth}
			\includegraphics[width=\linewidth]{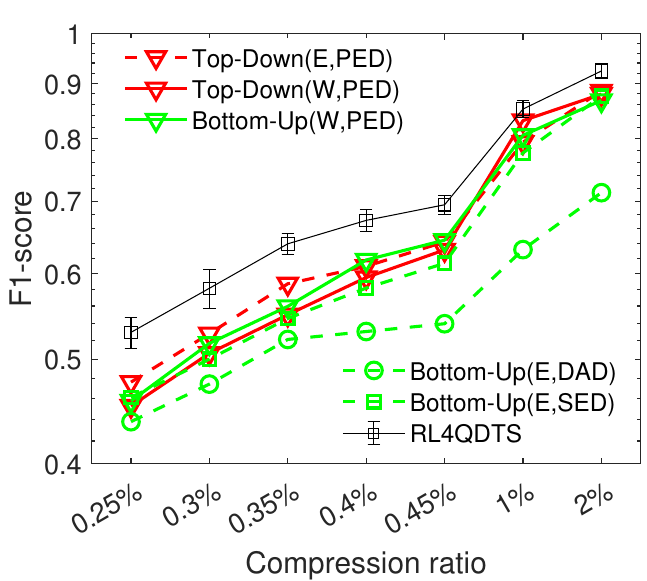}
		\end{minipage}\hspace{-3mm}
		&
		\begin{minipage}{0.2\linewidth}
			\includegraphics[width=\linewidth]{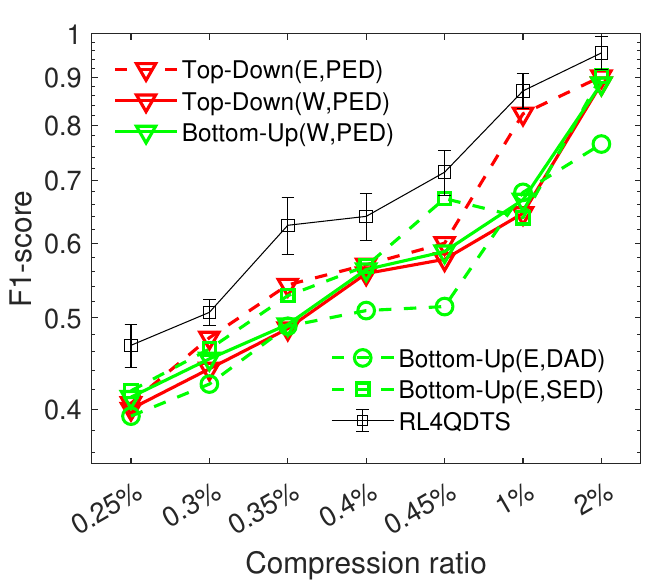}
		\end{minipage}
		\\
		(a) Range Query
		&
		(b) $k$NN Query (EDR)
		&
		(c) $k$NN Query (t2vec)
		&
		(d) Similarity Query
		&
		(e) Clustering
		\\
        \hspace{-3mm}
		\begin{minipage}{0.2\linewidth}
			\includegraphics[width=\linewidth]{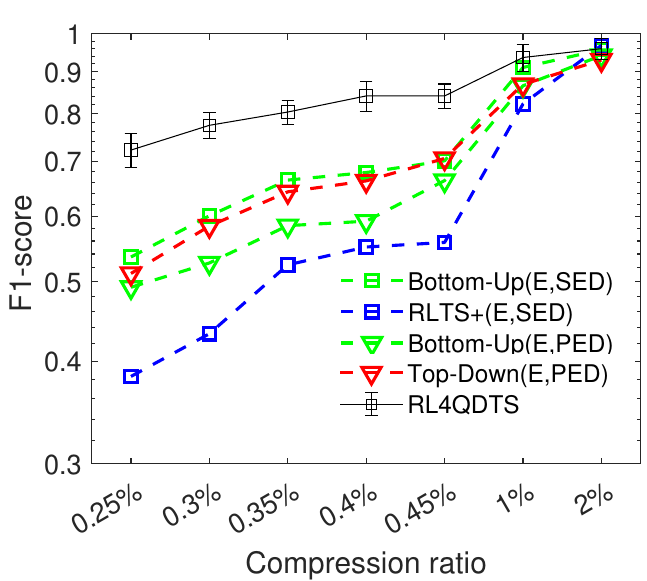}
		\end{minipage}\hspace{-3mm}
		&
		\begin{minipage}{0.2\linewidth}
			\includegraphics[width=\linewidth]{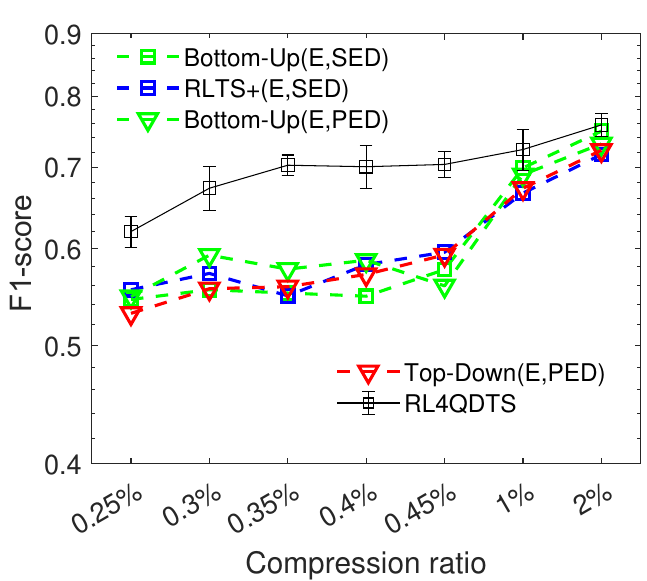}
		\end{minipage}\hspace{-3mm}
		&
		\begin{minipage}{0.2\linewidth}
			\includegraphics[width=\linewidth]{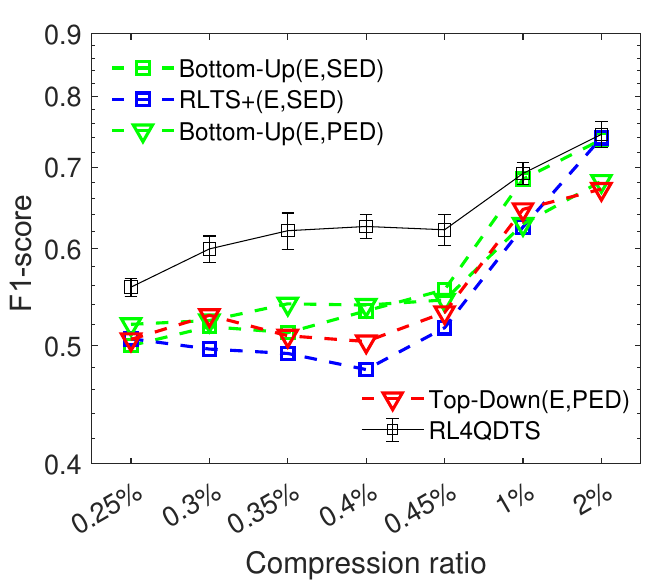}
		\end{minipage}\hspace{-3mm}
		&
		\begin{minipage}{0.2\linewidth}
			\includegraphics[width=\linewidth]{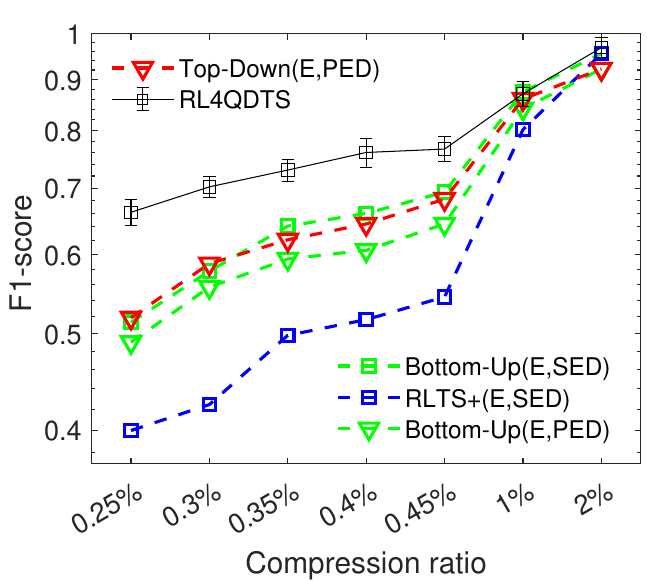}
		\end{minipage}\hspace{-3mm}
		&
		\begin{minipage}{0.2\linewidth}
			\includegraphics[width=\linewidth]{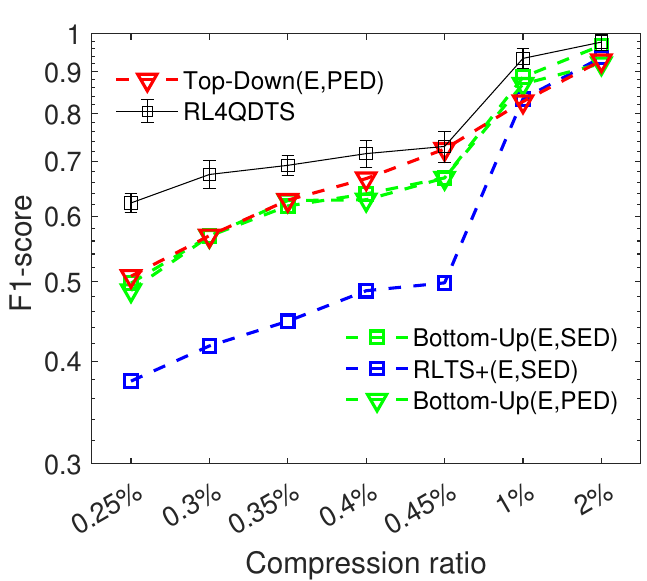}
		\end{minipage}
		\\
		(f) Range Query
		&
		(g) $k$NN Query (EDR)
		&
		(h) $k$NN Query (t2vec)
		&
		(i) Similarity Query
		&
		(j) Clustering
	\end{tabular}
 	\vspace*{-4mm}
	\caption{{\zhengII{Comparison with skylines on Geolife (data distribution (a)-(e) and Gaussian distribution (f)-(j)).
 }}}\label{fig:comparison_ap_geolife}
	\vspace*{-5mm}
\end{figure*}

\begin{figure*}[]
	\centering
	\hspace*{-4mm}
	\begin{tabular}{c c c c c}
        \hspace{-3mm}
		\begin{minipage}{0.2\linewidth}
			\includegraphics[width=\linewidth]{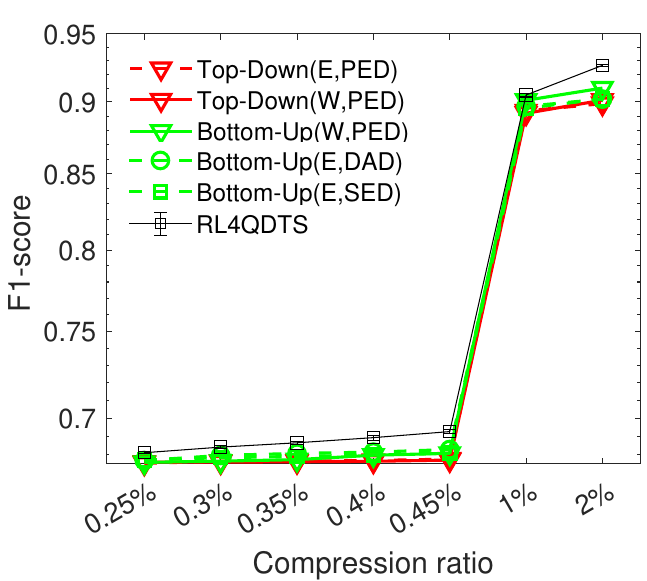}
		\end{minipage}\hspace{-3mm}
		&
		\begin{minipage}{0.2\linewidth}
			\includegraphics[width=\linewidth]{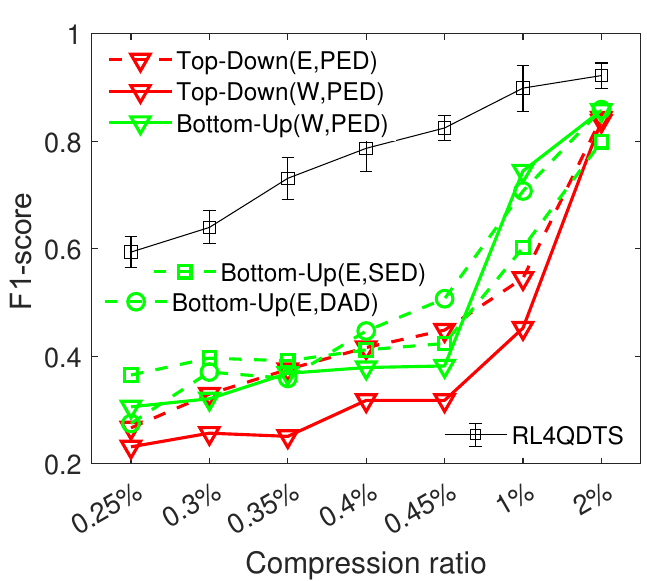}
		\end{minipage}\hspace{-3mm}
		&
		\begin{minipage}{0.2\linewidth}
			\includegraphics[width=\linewidth]{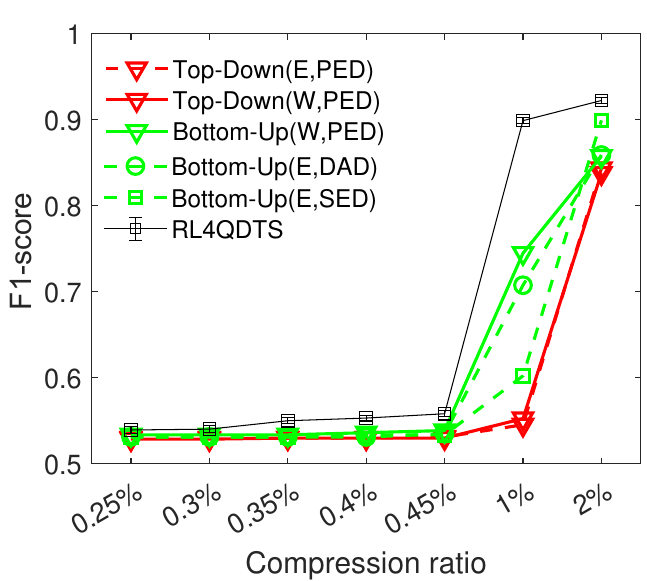}
		\end{minipage}\hspace{-3mm}
		&
		\begin{minipage}{0.2\linewidth}
			\includegraphics[width=\linewidth]{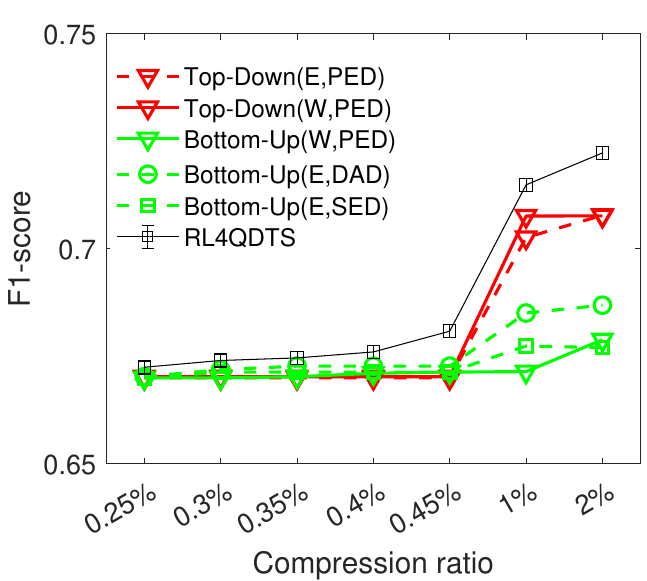}
		\end{minipage}\hspace{-3mm}
		&
		\begin{minipage}{0.2\linewidth}
			\includegraphics[width=\linewidth]{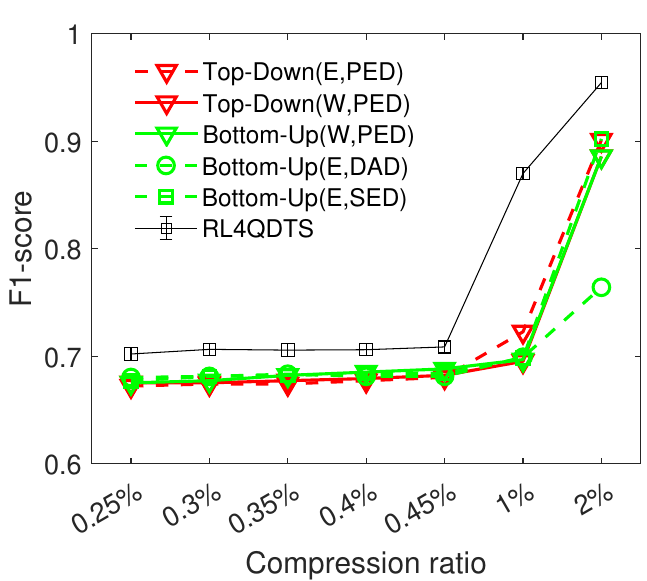}
		\end{minipage}
		\\
		(a) Range Query
		&
		(b) $k$NN Query (EDR)
		&
		(c) $k$NN Query (t2vec)
		&
		(d) Similarity Query
		&
		(e) Clustering
		\\
        \hspace{-3mm}
		\begin{minipage}{0.2\linewidth}
			\includegraphics[width=\linewidth]{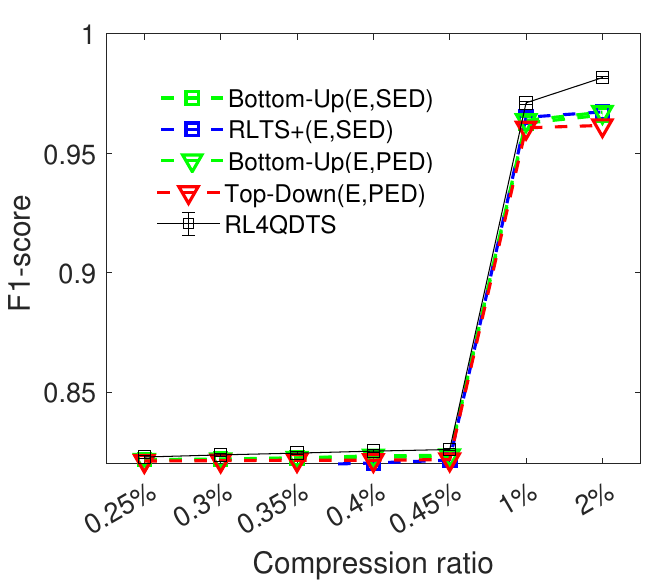}
		\end{minipage}\hspace{-3mm}
		&
		\begin{minipage}{0.2\linewidth}
			\includegraphics[width=\linewidth]{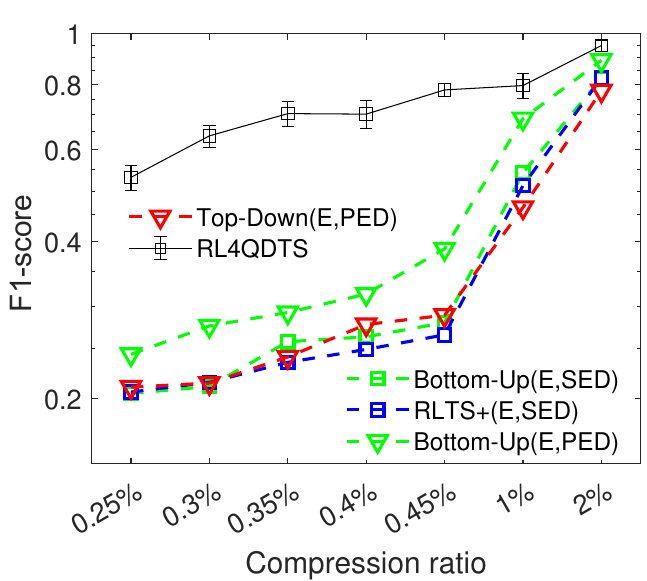}
		\end{minipage}\hspace{-3mm}
		&
		\begin{minipage}{0.2\linewidth}
			\includegraphics[width=\linewidth]{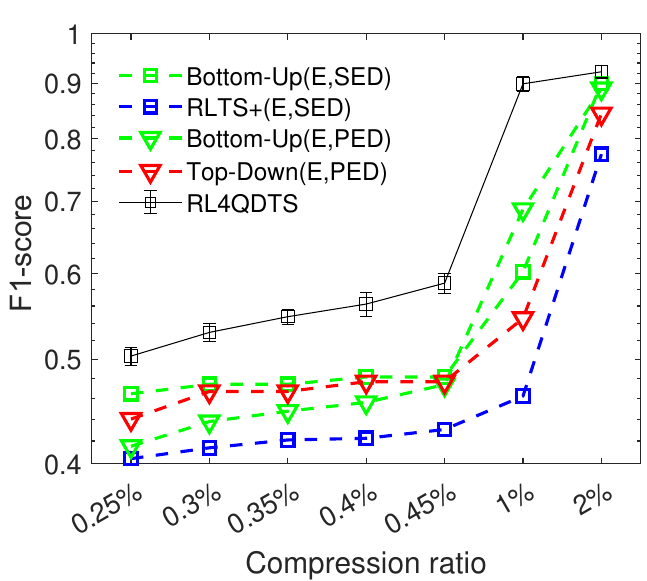}
		\end{minipage}\hspace{-3mm}
		&
		\begin{minipage}{0.2\linewidth}
			\includegraphics[width=\linewidth]{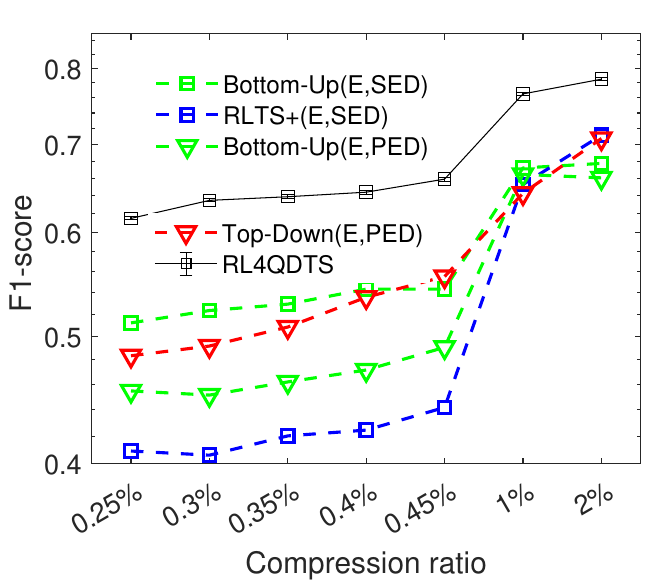}
		\end{minipage}\hspace{-3mm}
		&
		\begin{minipage}{0.2\linewidth}
			\includegraphics[width=\linewidth]{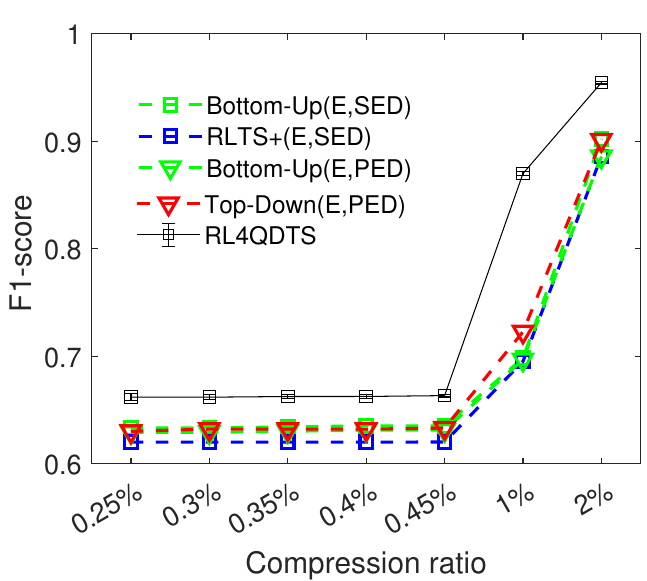}
		\end{minipage}
		\\
		(f) Range Query
		&
		(g) $k$NN Query (EDR)
		&
		(h) $k$NN Query (t2vec)
		&
		(i) Similarity Query
		&
		(j) Clustering
	\end{tabular}
	\vspace*{-3mm}
	\caption{{\zhengII{Comparison with skylines on T-Drive (data distribution (a)-(e) and Gaussian distribution (f)-(j)).}}}
	\label{fig:comparison_ap_tdrive}
	\vspace{-6mm}
\end{figure*}


\begin{figure*}[]
	\centering
	\hspace*{-4mm}
	\begin{tabular}{c c c c c}
        \hspace{-3mm}
		\begin{minipage}{0.2\linewidth}
			\includegraphics[width=\linewidth]{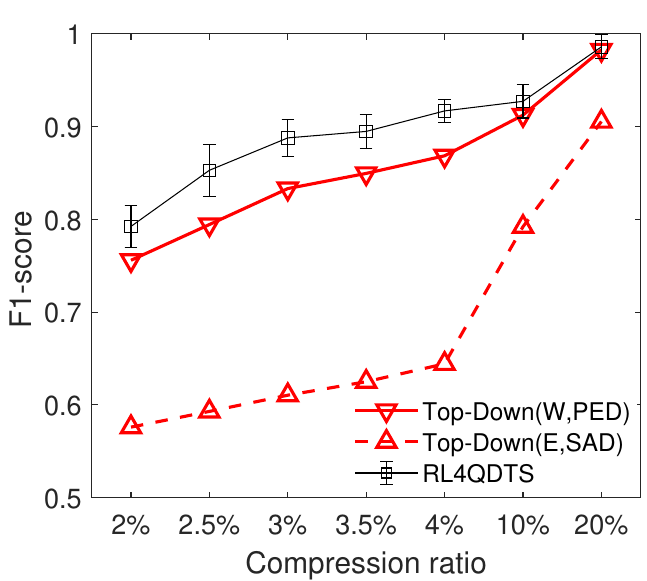}
		\end{minipage}\hspace{-3mm}
		&
		\begin{minipage}{0.2\linewidth}
			\includegraphics[width=\linewidth]{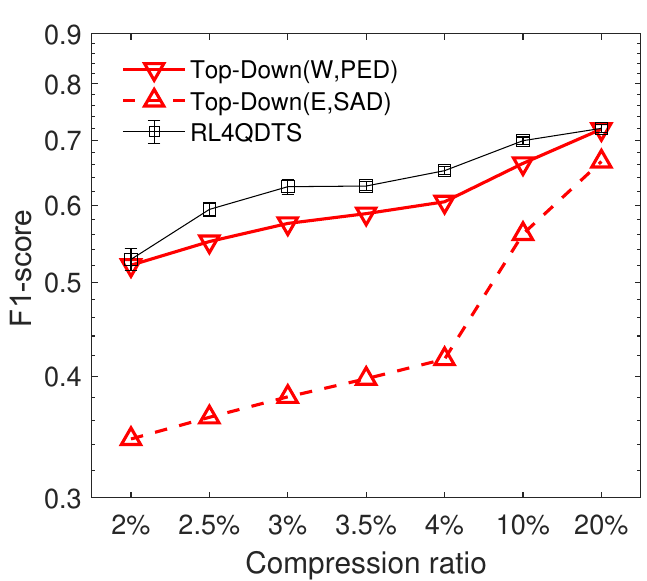}
		\end{minipage}\hspace{-3mm}
		&
		\begin{minipage}{0.2\linewidth}
			\includegraphics[width=\linewidth]{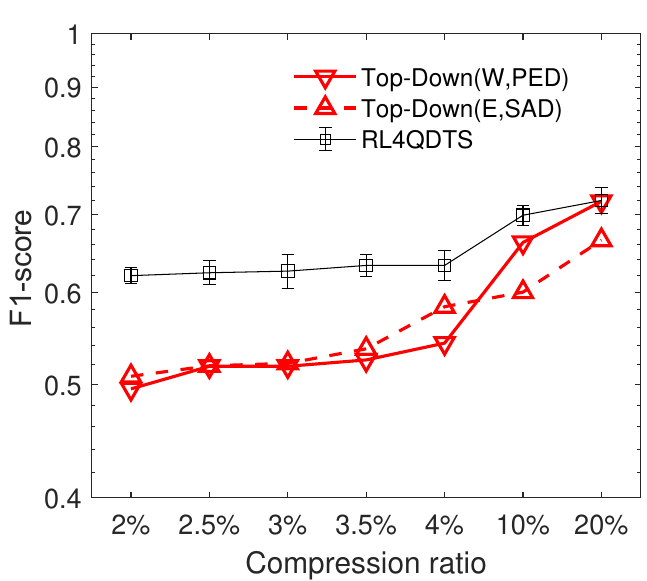}
		\end{minipage}\hspace{-3mm}
		&
		\begin{minipage}{0.2\linewidth}
			\includegraphics[width=\linewidth]{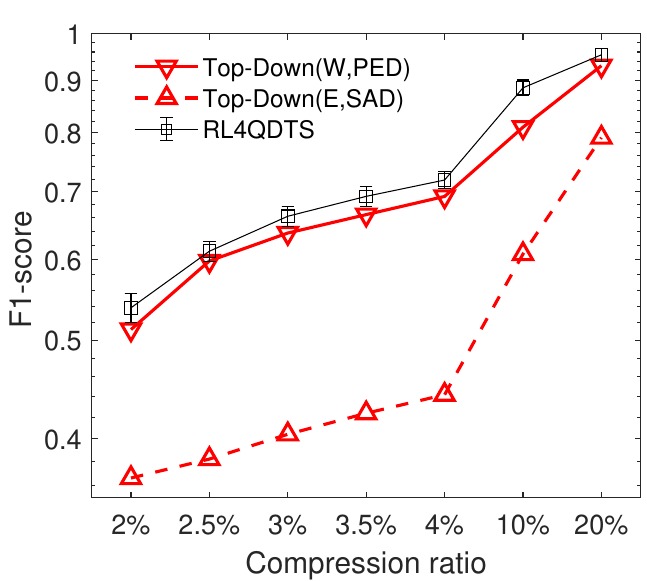}
		\end{minipage}\hspace{-3mm}
		&
		\begin{minipage}{0.2\linewidth}
			\includegraphics[width=\linewidth]{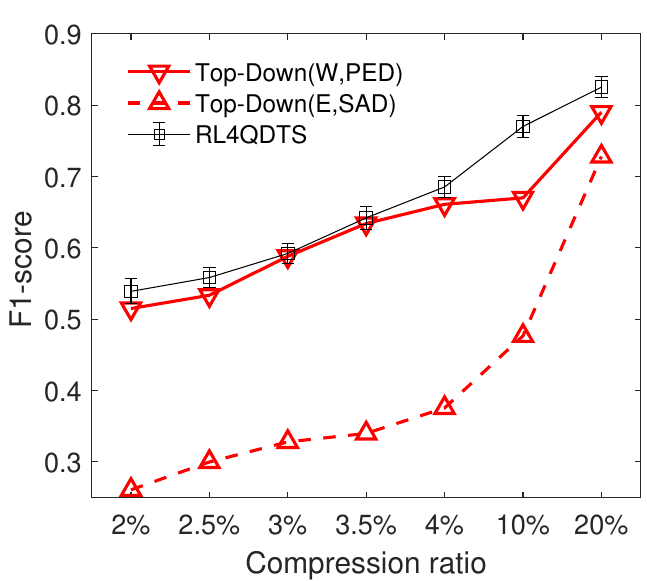}
		\end{minipage}
		\\
		(a) Range Query
		&
		(b) $k$NN Query (EDR)
		&
		(c) $k$NN Query (t2vec)
		&
		(d) Similarity Query
		&
		(e) Clustering
	\end{tabular}
	\vspace*{-4mm}
	\caption{{\zhengII{Comparison with skylines on Chengdu (real distribution (a)-(e)).}}}
	\label{fig:comparison_ap_chengdu}
	\vspace*{-7mm}
\end{figure*}


\subsection{Experimental Results}
\label{sec:results}

\noindent\textbf{(1) Effectiveness evaluation (skyline selection of existing algorithms).} 
%
To enable targeted comparisons, we select the skylines of the 25 baselines for each query task. Using a trajectory database $D$ with approximately 1.5 million points, we set the storage budget for simplification to $W=0.25\% \cdot N$ for Geolife (or $W=2\% \cdot N$ for Chengdu). Figure~\ref{fig:skyline} presents the effectiveness of the algorithms for five query tasks across three query distributions. 
For each task, we query 100 times and report the average results of the $F_1$-score as described in Section~\ref{subsec:problem}. We conclude the selected baselines for comparisons as follows.
For the \underline{data distribution}, Top-Down(E,PED), Top-Down(W,PED), Bottom-Up(W,PED), Bottom-Up(E,DAD), and Bottom-Up(E,SED) are on the skyline. 
For the \underline{Gaussian distribution}, Bottom-Up(E,SED), RLTS+(E,SED), Bottom-Up(E,PED), and Top-Down(E,PED) are on the skyline. 
For the \underline{real distribution}, Top-Down(W,PED) and Top-Down(E,SAD) are on the skyline.

\smallskip\noindent
{\zhengb \textbf{(2) Effectiveness evaluation (comparison with skyline).} 
We compare \texttt{RL4QDTS} with the selected skyline methods for each query task. We vary the storage budget $W$ from $0.25\% \cdot N$ to $2\% \cdot N$ for Geolife and T-Drive, and $2\% \cdot N$ to $20\% \cdot N$ for Chengdu. Here, a compression ratio of 0.25\% means that we reduce the data by a factor of 400 (=100/0.25). That is, the lower the compression ratio is, the more the data is reduced. With the current settings of the compression ratio, (1) the data reduction rate is some 50-400 times for Geolife and T-Drive and 5-50 times for Chengdu, which looks reasonable in practice, and (2) the accuracy of the query processing is also acceptable (e.g., the F1 score is at least 60\% in many cases as shown later on). We note that the trajectories in the Chengdu dataset are shorter than those in other datasets, as shown in Table~\ref{tab:dataset}, and thus we set the budget higher for Chengdu. Figure~\ref{fig:comparison_ap_geolife} shows the results on the two query distributions (i.e., data and Gaussian) on Geolife. For \texttt{RL4QDTS}, we show its error bars obtained by running the algorithm 50 times as described in Section~\ref{sec:setup}. The results based on T-Drive and Chengdu, shown in Figure~\ref{fig:comparison_ap_tdrive} and Figure~\ref{fig:comparison_ap_chengdu}, respectively, demonstrate trends similar to those seen on Geolife. 
Overall, \texttt{RL4QDTS} consistently outperforms existing error-driven methods across different budgets, query tasks, generation distributions, and real datasets. This is because \texttt{RL4QDTS} aims to preserve query quality directly, while existing methods minimize a given error measure without considering query quality directly.}

\begin{table}[t]
\setlength{\tabcolsep}{8pt}
\centering
\caption{Ablation study for \texttt{RL4QDTS} (Geolife).}
\vspace*{-3mm}
\begin{tabular}{|c|c|c|}
\hline
Effectiveness           & Range Query & Time (s)  \\ \hline
\texttt{RL4QDTS}           &\bm{$0.733\pm 0.018$} &61.11         \\ \hline
w/o Agent-Cube &$0.673\pm0.023$ &50.32        \\ \hline
w/o Agent-Point &$0.716\pm0.021$ &59.31         \\ \hline
w/o Agent-Cube and Agent-Point &$0.641\pm0.023$ &48.18   \\ \hline
\end{tabular}
\label{tab:ablation}
\vspace*{-7mm}
\end{table}

\begin{figure}[t]
	\centering
	\hspace{-5mm}
	\begin{tabular}{c c}
        \begin{minipage}{0.47\linewidth}
			\includegraphics[width=\linewidth]{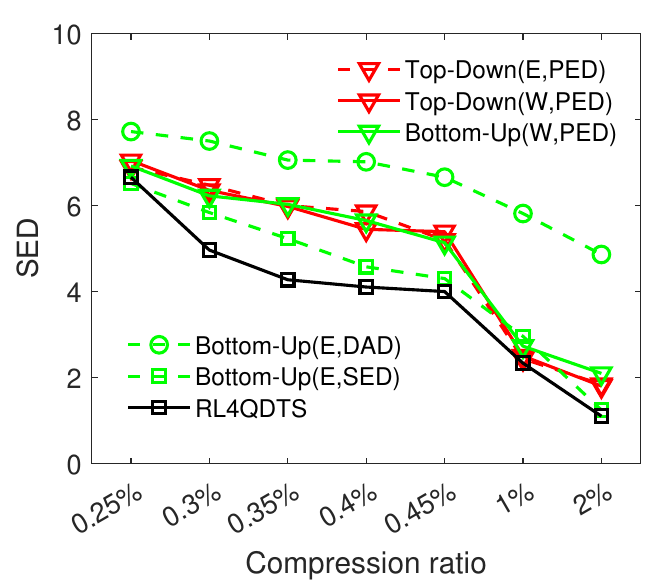}
		\end{minipage}\hspace{-2mm}
		&
		\begin{minipage}{0.47\linewidth}
			\includegraphics[width=\linewidth]{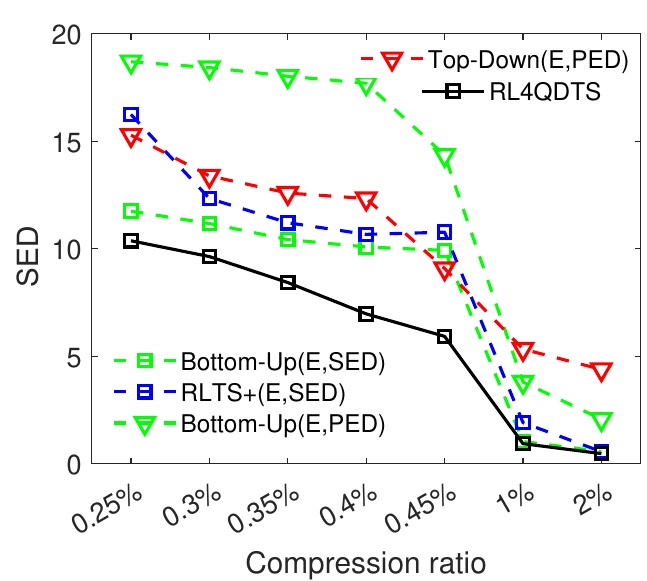}
		\end{minipage}\hspace{-2mm}
		\\
		(a) 
        {\chengrr Data distribution}
		&
        {\chengrr (b) Gaussian distribution}
        \\
	\end{tabular}
	\vspace*{-2mm}
	\caption{{\zhengII{Deformation study {\chengrr (SED errors of trajectories returned by queries).}}}}
	\label{fig:deformation}
	\vspace*{-3mm}
\end{figure}

\smallskip\noindent
{\zheng{\textbf{(3) Effectiveness evaluation (ablation study).} We conduct an ablation study to investigate the effects of Agent-Cube and Agent-Point in \texttt{RL4QDTS}. (1) We drop Agent-Cube by setting the start level $S=9$ and the end level $E=10$, so that Agent-Cube reduces to randomly sampling a cube according to the data distribution and then returning the cube to Agent-Point. (2) We drop Agent-Point and instead insert the point with the maximum value into a simplified database. (3) We drop both Agent-Cube and Agent-Point with the strategies described above.
Table~\ref{tab:ablation} reports the average results of 100 range queries with a distribution that follows the data distribution on a randomly sampled trajectory database with around 1.5 million points from Geolife. Overall, all components contribute to the result,
and the two agents cooperate to optimize query quality.
%
}}



\smallskip
\noindent
{\zhengII{\textbf{(4) Effectiveness evaluation (deformation study).} We study the deformation of the trajectories {\chengrr returned by} queries. In Figure~\ref{fig:deformation}, we run 100 range queries {\chengr with the} data and Gaussian distributions, and report the average SED of the returned trajectories, which measures the deformation in terms of SED between the original trajectories and their simplified ones. As expected, \texttt{RL4QDTS} is consistently lower than skyline methods, because \texttt{RL4QDTS} is a query-aware solution, which preserves more points for those trajectories to answer the queries. For the skyline methods, they fail to preserve the trajectories {\chengrr returned by} queries, though the distances (e.g., SED) can be optimized explicitly {\chengr for all trajectories including both the returned ones and others}.
}}

\smallskip\noindent {\zhengb \textbf{(5-8) Parameter study (varying parameters $S$ and $E$ in Agent-Cube, $K$ in Agent-Point, and $k$ in $k$NN query).} We evaluate the effect of (1) the start level $S$ and (2) the end level $E$ in Agent-Cube, (3) parameter $K$ that controls the state space of Agent-Point for decision-making, and (4) different $k$ in $k$NN queries with EDR and t2vec on Geolife. Overall, we observe that (1) a moderate setting of $S=9$ and $E=12$ brings the best effectiveness, (2) $K=2$ provides a reasonable trade-off between effectiveness and efficiency, and (3) the effectiveness improves as $k$ increases. The results are included in the technical report~\cite{TR} due to the page limit.
}

\begin{figure}[t]
	\centering
	\hspace{-3mm}
	\begin{tabular}{c c}
       \begin{minipage}{4cm}
        \includegraphics[width=6.5cm]{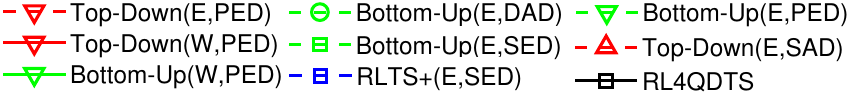}
       \end{minipage}
      \\
		\begin{minipage}{0.45\linewidth}
			\includegraphics[width=\linewidth]{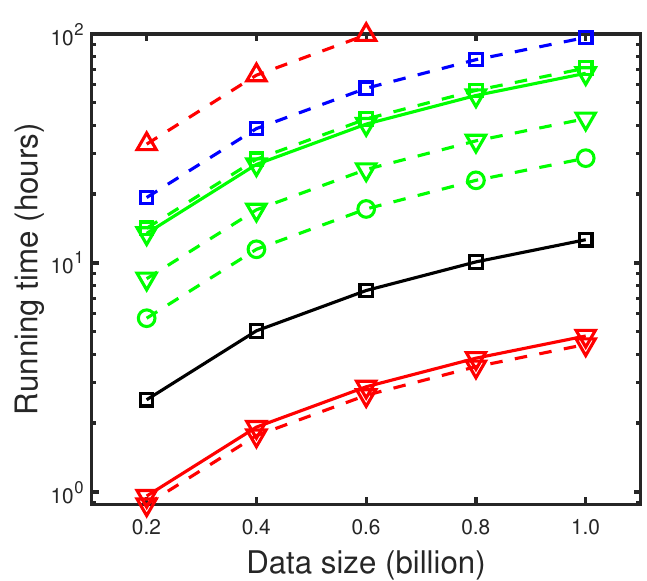}
		\end{minipage}\hspace{-2mm}
		&
		\begin{minipage}{0.45\linewidth}
			\includegraphics[width=\linewidth]{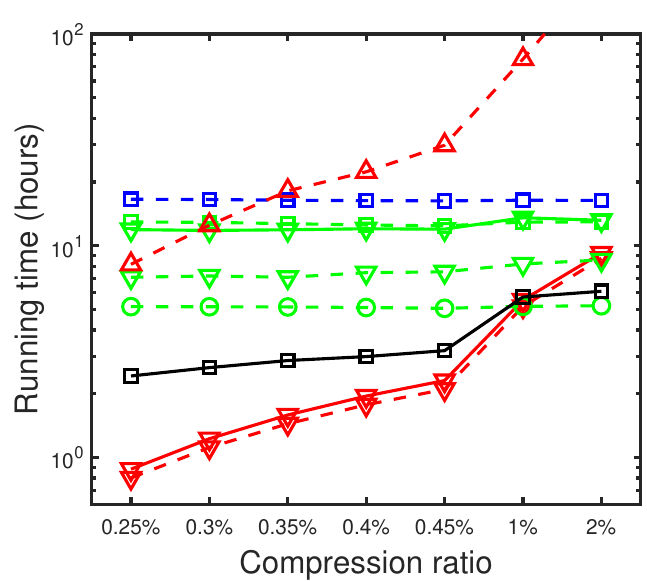}
		\end{minipage}\hspace{-2mm}
		\\
		(a) Varying data size
		&
		(b) Varying budget size
        \\
	\end{tabular}
	\vspace*{-2mm}
	\caption{{\zhengII{Efficiency evaluation (OSM).}}}
	\label{fig:efficiency}
	\vspace*{-7mm}
\end{figure}

\smallskip
\noindent
{\zheng{\textbf{(9) Scalability test (varying the data size $N$).} We study scalability when varying the database size on OSM. We compare all skyline methods as shown in Figure~\ref{fig:skyline}, and vary the trajectory database size $N$ from 0.2 billion to 1 billion points, with a fixed storage budget $W=0.25\% \cdot N$. The running times (the maximum is set to 100 hours) are shown in Figure~\ref{fig:efficiency}(a). 
%
%
Overall, \texttt{RL4QDTS} is faster than most existing methods, except for adaptations from Top-Down approaches. \texttt{RL4QDTS} enhances effectiveness through learned policies, incurring time costs for state construction and action sampling. 
%
%
{\zhengIII{The results are similar on other datasets and omitted.}}
%
}}

\smallskip
\noindent\textbf{(10) Efficiency evaluation (varying the budget size $W$).} We further study the effect of budget size $W$ from $0.25\% \cdot N$ to $2\% \cdot N$, with a fixed $N$ of 0.1 billion points. Figure~\ref{fig:efficiency}(b) illustrates the running time on Geolife. \texttt{RL4QDTS} is slower than the Top-Down adaptions, but is faster than the Bottom-Up adaptions by at least a factor of two times. {\zhengII{As $W$ increases, \texttt{RL4QDTS} becomes faster than Top-Down adaptions, because it computes the values based on a partial trajectory within a cube by Agent-Point; however, Top-Down adaptions computes that values based on a whole trajectory.
}}
%

\smallskip\noindent {\zhengb \textbf{(11) Training time.} We show the training time of \texttt{RL4QDTS} with (1) the number of trajectories and (2) parameter $\Delta$ on Geolife. Overall, we observe that (1) the setting of 6,000 trajectories is enough to obtain a good model with a reasonable training cost, and (2) a moderate $\Delta = 50$ is with the best effectiveness. The results are included in~\cite{TR}. 
}

\begin{figure}[t]
\centering
    \begin{minipage}{0.33\linewidth}
      \includegraphics[width=\linewidth]{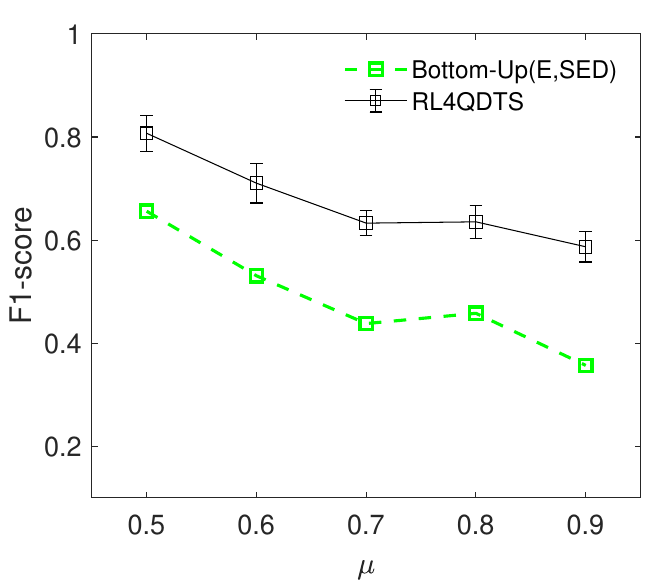}
    \end{minipage}
    \centering
    \begin{minipage}{0.33\linewidth}
      \includegraphics[width=\linewidth]{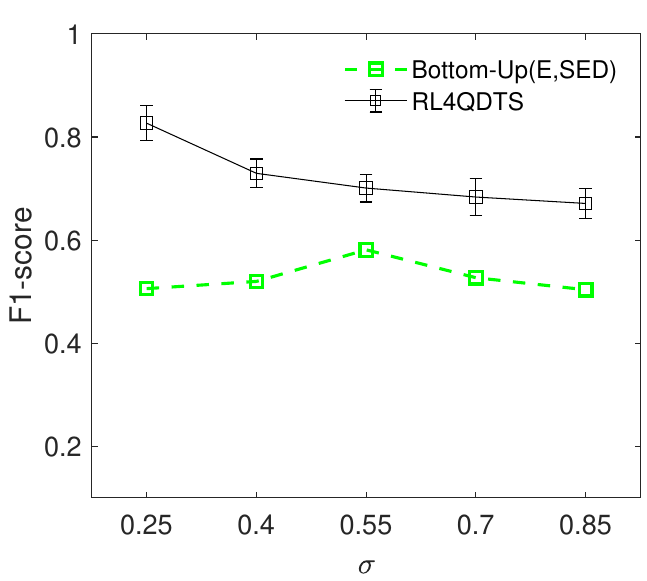}
    \end{minipage}
    \centering
    \begin{minipage}{0.33\linewidth}
      \includegraphics[width=\linewidth]{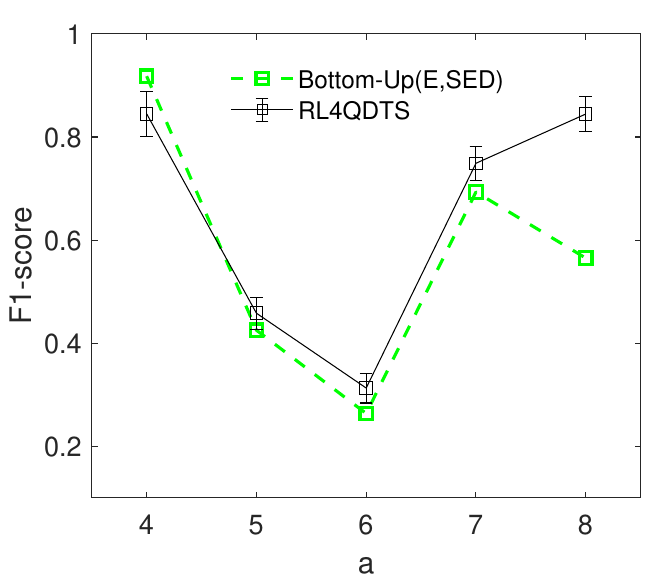}
    \end{minipage}
    \centering

    (a) Gaussian $\mu$ \qquad\quad (b) Gaussian $\sigma$ \qquad\quad (c) Zipf $a$
    
    \begin{minipage}{0.4\linewidth}
      \includegraphics[width=\linewidth]{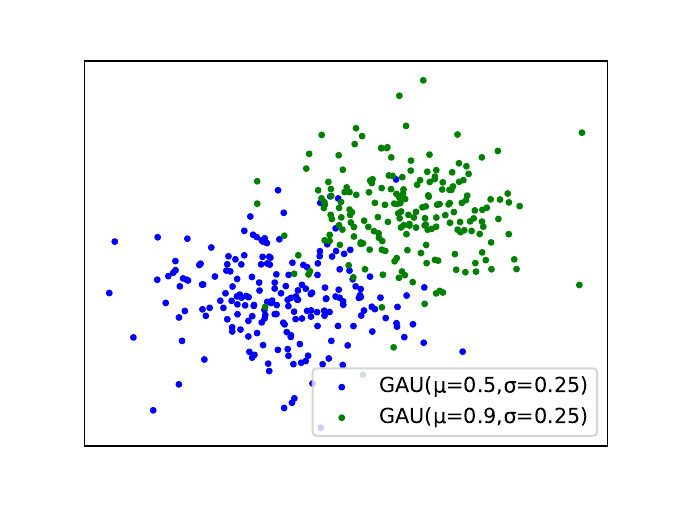}
    \end{minipage}\hspace{2mm}
    \centering
    \begin{minipage}{0.4\linewidth}
      \includegraphics[width=\linewidth]{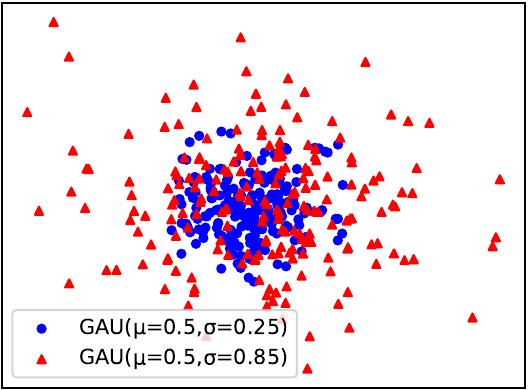}
    \end{minipage}
    
    (d) {\chengr Gaussian ($\mu = 0.9$) \quad (e) Gaussian ($\sigma = 0.85$)}

    \begin{minipage}{0.4\linewidth}
      \includegraphics[width=\linewidth]{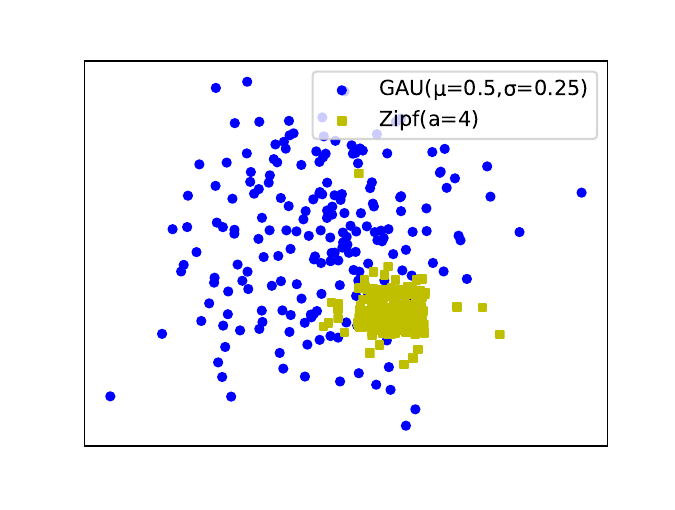}
    \end{minipage}\hspace{2mm}
    \centering
    \begin{minipage}{0.4\linewidth}
      \includegraphics[width=\linewidth]{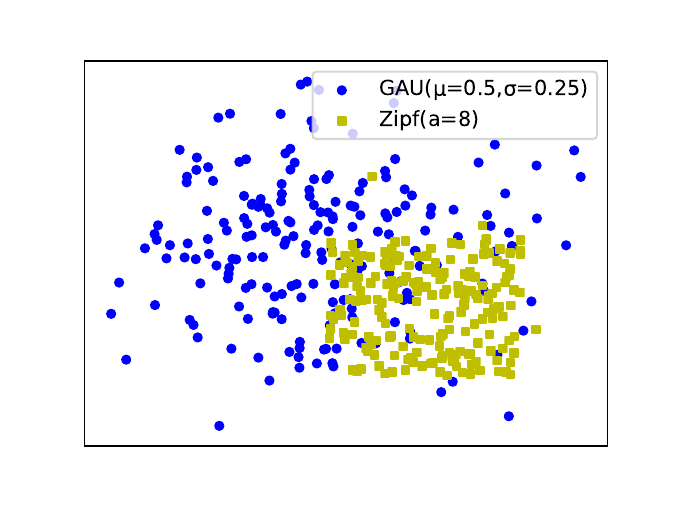}
    \end{minipage}
    
    {\chengr (f) Zipf ($a=4$) \qquad\quad (g) Zipf ($a=8$)}
\vspace*{-3mm}
\caption{Transferability test.}
\label{fig:transfer}
\vspace*{-7mm}
\end{figure}

\smallskip
{\zhengII{
\noindent\textbf{(12) Transferability test (with distribution changes).} 
We test \texttt{RL4QDTS} transferability in two scenarios. First, we train \texttt{RL4QDTS} with range queries following the Gaussian distribution ($\mu=0.5$ and $\sigma=0.25$) on Geolife and evaluate its effectiveness for range queries with varying $\mu$ (0.5 to 0.9) and $\sigma$ (0.25 to 0.85) (Figure~\ref{fig:transfer}(a) and (b)). This tests transferability for moderate query distribution changes. Second, using the same model trained with the Gaussian distribution, we test its effectiveness for range queries following a Zipf distribution with different exponent parameters $a$ (4 to 8) to assess transferability under significant distribution changes (Figure~\ref{fig:transfer}(c)). \texttt{RL4QDTS} consistently outperforms the baseline across all $\mu$ and $\sigma$ settings (Figure~\ref{fig:transfer}(a) and (b)). In Figure~\ref{fig:transfer}(c), \texttt{RL4QDTS} performs comparably well or better than the baseline despite drastic query distribution changes, demonstrating its robustness. This may be explained by that \texttt{RL4QDTS} does not rely on any error measures and preserves patterns and knowledge embedded in the data via neural networks, which remain useful to optimize queries even if the distributions are changed. We also visualize the distribution changes in Figure~\ref{fig:transfer}(d)-(g).

\section{CONCLUSION}
\label{sec:con}
{\zhengb We introduce the query accuracy-driven trajectory simplification problem, aiming to minimize the difference between query results on the original database and {\chengb on} a simplified one. Our novel solution, \texttt{RL4QDTS}, employs multi-agent reinforcement learning, enabling collective trajectory simplification while directly optimizing the QDTS problem's objective. Extensive experiments on real-world trajectory datasets demonstrate that \texttt{RL4QDTS} consistently outperforms existing EDTS algorithms in five query processing operations. One promising research direction is to explore data-driven approaches for compressing road network-based trajectories to enhance effectiveness and efficiency.}

\smallskip
{\CR 
\noindent\textbf{Acknowledgments:}
This research/project is supported by the National Research Foundation, Singapore under its AI Singapore Programme (AISG Award No: AISG-PhD/2021-08-024[T] and AISG Award No: AISG2-TC-2021-001); the Ministry of Education, Singapore, under its Academic Research Fund (Tier 2 Awards MOE-T2EP20220-0011 and MOE-T2EP20221-0013); and the Innovation Fund Denmark center, DIREC. Any opinions, findings and conclusions or recommendations expressed in this material are those of the author(s) and do not reflect the views of National Research Foundation, Singapore and Ministry of Education, Singapore.}

\bibliography{ref}
\bibliographystyle{IEEEtran}

\end{document}